\documentclass{aa}  
\usepackage[english]{babel}
\usepackage{amsmath}
\usepackage{graphicx}
\usepackage{natbib}
\usepackage[colorlinks=true, allcolors=blue]{hyperref}
\usepackage{txfonts}
\begin{document}

   \title{Modeling time delays from two reprocessors in active galactic nuclei}

   \author{Vikram Kumar Jaiswal
          \inst{1}
          \and
          Raj Prince\inst{1}
          \and
          Swayamtrupta Panda\inst{2,1}\thanks{CNPq Fellow}
          \and
          Bo\.zena Czerny\inst{1}
          }

   \institute{Center for Theoretical Physics, Polish Academy of Sciences,
   Al. Lotnik\'ow 32/46, 02-668 Warsaw, Poland
   \and Laborat\'orio Nacional de Astrof\'isica - MCTIC, R. dos Estados Unidos, 154 - Na\c{c}\~oes, Itajub\'a - MG, 37504-364, Brazil
             }

   \date{Received ???; accepted ???}
 
  \abstract  
   {Continuum time delays from accretion disks in active galactic nuclei (AGNs) have long been proposed as a tool for measuring distances to monitored sources. However, the method faces serious problems as a number of effects must be taken into account, including the contribution from the broad line region (BLR).}
   {In this paper, we model the expected time delays when both the disk reprocessing of the incident X-ray flux and further reprocessing by the BLR are included, with the aim to see whether the two effects can be disentangled.}
   {We used a simple response function for the accretion disk, without relativistic effects, and we used a parametric description to account for the BLR contribution. We included only the scattering of the disk emission by the BLR inter-cloud medium. We also used artificial light curves with one-day samplings to check whether the effects are likely to be seen in real data.}
   {We show that the effect of the BLR scattering on the predicted time delay is very similar to the effect of the rising height of the X-ray source, without any BLR contribution. This brings additional degeneracy for potential applications in the future, when attempting to recover the parameters of the system from the observed time delays in a specific object. Both effects, however, modify the slope of the delay-versus-wavelength curve when plotted in log space, which opens a way to obtaining bare disk time delay needed for cosmology. In addition, when the disk irradiation is strong, the modification of the predicted delay by the BLR scattering and by X-ray source height become considerably different. In this regard, simulations of the expected bias are also presented.}
   {}


   \maketitle
%________________________________________________________________

\section{Introduction}

The standard  $\Lambda$CDM cosmological model is currently under vigorous discussion and testing via a state-of-the-art approach based on current and upcoming astronomical instruments \citep{2017NatAs...1E.121F}. This global model describes the evolution of the entire Universe, so the measurements of the global model parameters should give the same values independently of how and where they are measured. One such parameter is the current (i.e., at redshift zero) expansion rate of the Universe, the Hubble constant, $H_0$. However, the local direct measurements based on SN Ia, calibrated predominantly with Cepheid stars, give average values of $H_0$ of the order of $74\,{\rm km\, s^{-1} Mpc^{-1}}$ (e.g., $H_0 = 74.03 \pm 1.42\,{\rm km\, s^{-1} Mpc^{-1}}$, \citealt{riess2019}; $H_0 = 73.2 \pm 1.3\,{\rm km\, s^{-1} Mpc^{-1}}$, \citealt{riess2021}). At the same time, measurements of the cosmic microwave background (CMB) properties imply $67.4 \pm 0.5\,{\rm km\, s^{-1} Mpc^{-1}}$ \citep{Planck2020} when the standard $\Lambda$CDM is used to derive $H_0$. In the observational cosmology this discrepancy in the Hubble constant is known as Hubble Tension.
 The detection of Hubble tension suggests a need of different cosmological model to explain the local universe. However, before going into detail of deriving a different cosmological model, we have to be very sure that the tension really exists -- hence, a range of different independent methods is required to probe it.
A similar tension between the Planck results and the local measurements shows up for most (and quite numerous) methods \citep[for the most recent comprehensive review, see][in particular Fig. 1 therein]{eleonora2021}. However, the disagreement is not at all that clear since the systematic errors in each method are difficult to assess.

Therefore, a relatively simple and direct method, which would not require the involvement of the distance ladder would be extremely useful to fix the problem. One such method is the one based on continuum time delays in accretion disks in active galactic nuclei (AGNs), proposed by \citet{collier1999} (see also \citealt{oknyanskij1999} for a torus-based version of the idea). The method is effectively based on measuring the size of the accretion disk at different wavelengths and comparing it with the classical accretion disk model of \citet{SS1973}. This comparison, due to a specific scaling of both the monochromatic flux and the effective temperature with the product of the black hole mass and accretion rate \citep[see e.g.,][]{2018ApJ...866..115P}, allows us to get the distance to the source directly, from the measured time delay $\tau$ between the two wavelengths and the observed monochromatic flux, $f_{\nu}$, at one of these wavelengths, without any hidden dependence on the black hole mass and accretion rate. The time delay, $\tau$, as predicted by the theory, depends on the wavelength $\lambda$ as $\tau \propto \lambda^{4/3}$, and the proportionality coefficient is also strictly predicted by the theory. This coefficient contains the observed flux and the distance, thus offering the possibility to obtain the redshift-independent distance to the source knowing the time delays and observed fluxes.

Observational monitoring of several sources confirmed the expected delay pattern \citep[e.g.,][]{collier1999,cackett2007,pozo2019,lobban2020}, specifically, the proportionality: $\tau \propto \lambda^{4/3}$. However, the observationally determined proportionality constant was frequently deemed as being much too high (by some 40\%  up to a factor of a few) in comparison with the theory, that is, when the standard cosmology was used \citep[e.g.,][]{collier1999,cackett2007,lobban2020,GuoW2022} and the narrow filters used by \citet{pozo2019} did not help. Some sources have found a good agreement with expectations, particularly if the height of the irradiating source and/or the extended character of the reprocessor are included \citep[][]{kammoun2021_data}; some authors have found a disagreement only at a single wavelength band, close to the Balmer edge \citep[e.g.,][]{edelson2015,Kammoun2019,McHardy2018,cackett2018,hernandez2020,cackett2020}. Hints of a considerable problem with regard to the contribution from the more distant reprocessing region, namely, the broad line region (BLR), have appeared \citep{chelouche2019}. In his most recent paper, \citet{netzer2021} concluded that (for sources analysed by him) most of the reprocessing actually happens in the BLR region, making a continuum time delay longer than the predicted value from the disk itself. A similar suggestion was made earlier by \citet{lawther2018} for the case of NGC 5548. If this is true, it is very difficult to disentangle the BLR time delay with the continuum disk time delay -- hence, these objects cannot be used for the cosmological purpose. The source of contamination comes, apart from strong emission lines, also from broad-band spectral features such as the Fe {\sc ii} pseudo-continuum and the Balmer continuum \citep{wills1985}. The problems of reconciling the disk size with the standard model also appeared in microlensing studies \citep{rauch1991,mosquera2013}, but the presence of the additional reprocessing medium most likely solves this problem as well. 

In the present paper, we model the combined reprocessing by two media: an accretion disk and the  extended BLR, with the aim to find a way to disentangle efficiently these two effects. Finally, we aim to use these results to reconstruct the disk time delay. 

\section{Method}

We performed a set of numerical simulations that allows us to see whether the adopted geometry for the disk and the BLR region can be recovered in measurements of the time delays. We created artificial light curves to mimic the incident radiation, assumed a set of parameters describing the disk and BLR reprocessing, and, finally, calculated the time delays using interpolated cross-correlation function (ICCF) method to see whether the geometry can be determined and, in particular, the conditions under which the time delay related to the accretion disk alone can be recovered in such simulations.

\subsection{Incident light curve simulation}
\label{sect:curves}

According to the general picture used in the description of the disk reprocessing, we assume that variable X-ray emission is responsible for the variability of the disk emission \citep{rokaki1993}.

We modeled the lightcurve using the algorithm of \citet{TK1995} (hereafter, the TK method), which is based on the adopted shape of the power spectrum. 
We modeled the X-ray power spectrum as a broken power law, with two breaks and three slopes. The higher frequency break has been relatively well studied for a number of AGNs.
For a given set of source parameters, the relation between the black hole mass, the Eddington rate, and the position of the break is given by \citet{mchardy2006}. Older and newer measurements are roughly consistent with this law \citep[e.g.,][]{czerny1999,czerny2001,markowitz2010}. For the slopes, we assumed -2, -1, and 0 , respectively.  For the lower frequency break, we assumed that its timescale is by a factor of 100 longer than the short timescale break. This is somewhat arbitrary, since the long timescale trends are not well measured. Alternatively, we could use a bending power law as, for example, in \citet{georgakakis2021}.
The normalization of the power spectrum is then adjusted to the required level of the source variability. This model is certainly better than a damped random walk (DRW), corresponding to a single break and slopes -2 and 0, proposed by \citet{kelly2009} for modeling the optical variability of AGNs. As shown by \citet{Yu2022}, more  advanced models are needed for precise description of quasar variability in Stripe 82. For our purposes, TK method is satisfactory, as it broadens the frequency range in comparison to DRW.

\subsection{Accretion disk reprocessing}
\label{sect:disk}

Our description of the disk reprocessing is relatively simple. For the geometry, we assumed a simple lamppost model that represents the X-ray corona, namely, one that is frequently adopted for the disk reprocessing in compact X-ray binaries and AGNs \citep[e.g.,][]{Martocchia_Matt_1996, Miniutti_Fabian_2004, Niedzwiecki_etal_2016}, and the disk height is neglected. We do not include general relativity (GR) effects and we assume perfect thermalization of the incident radiation. Once the X-ray photon hits the disk all the radiation get absorbed by the disk which increases the disk temperature locally.  
Thus, in our model, we do not consider any energy-dependent reflection as in \citet{Kammoun2019} but all the incident emission is absorbed by the disk and gets reprocessed.

The response of the disk to the variable irradiation was recently studied in much more detail by \citet{Kammoun2019,kammoun2021}. Their approach included numerous effects, such as the assumption of the Kerr metric in the description of the disks, full GR treatment of the photon propagation, and energy-dependent reflection of the X-rays by the disk surface.   We do not aim to achieve such a detailed approach here. Instead,  we adopt the rather simple approach  of \citet{SS1973} for the disk and geometrical optics, where a perfect thermalization of the incident X-ray flux is considered. In creating our software, we aim to check qualitatively the effect of the second reprocessor on the measured time delay and we do it not only with the use of the transfer function approach, but also with the use of artificial light curves simulating the data cadence and quality. 
In such cases, we simply carry out a direct calculation of  a sequence of disk reactions to variable irradiation by X-ray flux, where the incident flux can vary with arbitrarily large amplitudes. 

Our numerical scheme is thus different from the one used by \citet{Kammoun2019,kammoun2021}. We first set the radial grid covering the accretion disk between $R_{isco}$ and $R_{out}$ by a variable sample size. The radial bin size is quasi-logarithmic, to allow for a proper resolution at the inner parts. Specifically, we define this grid by the following formula, $dR = 0.085*(\frac{R}{R_{\rm isco}})^{0.85}$, and for each "R," we also increase the grid step in angular ($\phi$) direction by $d\phi = \frac{1.5700}{N_{div}}$. For a given R and $\phi$, we calculate the Cartesian (x,y) coordinate in the disk plane, and the surface element, $ds = R*dR*d\phi*R_g*R_g$. The disk height is neglected, that is, we assume $z = 0.$
For a given x, y coordinate, we calculate the total delay $\tau_{total}(x,y)$.  This delay is the sum of time $\tau_d(r)$ taken by photon to reach the given disk location from corona, located at the height, $H,$ along the symmetry axis to accretion disk, and the time to reach observer after disk reprocessing, $\tau_{do}(x,y)$. For convenience, we define this last time delay with respect to the plane crossing the equatorial plane at $R_{out}$, $\phi = 0$, and perpendicular to the direction towards the observer. This delay depends on the inclination angle $i$, the corona height, the position on the accretion disk, and the black hole mass. All the delays are calculated with respect to the photon generated from the corona.  

Once we have the total delay $\tau(r)$ for a differential disk elements, we calculate the new temperature from the total flux, $F$, the sum of flux generated from the non-irradiated disk, and irradiation. We thus used the following expression:\  

\begin{equation}
F(r,t+\tau_d(r)) = \left(\frac{3GM\Dot{M}}{8 \pi r^3}\left(1-\sqrt{\frac{6}{r}}\right)\right)+\left(\frac{L_x(t)h}{4\pi r^3}\right).
\end{equation}

The local temperature at each moment is calculated from the local flux assuming a blackbody emission:
\begin{equation}
T_{eff}(r,t +\tau_d(r)) = \left[\left(\frac{3GM\Dot{M}}{8 \pi r^3 \sigma_B}\left(1-\sqrt{\frac{6}{r}}\right)\right)+\left(\frac{L_x(t)h}{4\pi r^3 \sigma_B}\right)\right]^{\frac{1}{4}},\\
\end{equation}
where $\sigma_B$ is the Boltzmann constant.
We do not include the color correction, which would affect the normalization of the time delay but does not affect the basic trends with the parameters that are the topic of our study. However, the inclusion of the color correction indeed makes all the delays longer, which is critical for actual data fitting.

From Planck's formula, we can generate the entire spectrum for the given differential area only to store it in a photon table, which is a 2D matrix \textbf{P(t',${\boldsymbol{\mathrm{\lambda}}}$)}, using a determined time delay appropriate for the disk position and wavelength. Every flux element of the photon table has a unique delay and a wavelength $\lambda$. We chose the values of $\lambda$ from 1000-10000 \AA~ for the simulation, using a logarithmic scale grid for selection. The time delay step is either linear or logarithmic, depending on the need.

We assume that the incident light curve is provided with equal bin size, $\Delta$t. We treat such a light curve as a histogram, which is a step function with  $\Delta$t. Thus, the time bin size of the photon table matches the resolution of the incident light curve and the time span must be long enough to cover the duration of the incident lightcurve, $t_{irrad}$. For each incident light curve bin, we include this uncertainty of $\Delta$t of the photons arrival when locating the flux contribution into the photon table. We first performed the loop with respect to incident light curve bins for a given location at the disk, and then repeated the process for all the elements of the disk surface, thus creating the lightcurve expected from the irradiated disk, equivalent to:
\begin{equation}
\label{eq:lum}
L_{disk}(\lambda,t) = {1 \over \Delta_t}\int^{t_{irrad}}_{t_{min}}\int_{S_{disk}} B_{\lambda}(T_{eff}(r,(t^{'}-\tau_{do}(x,y))ds dt^{'},
\end{equation}
for a very dense grid. 

We can calculate the response function of the disk, $\psi_d$, which is a very useful concept if the response of the medium is linear  \citep[see e.g., Equation 9 of][]{peterson1993}. In this case, we replace the light curve $L_X(t)$ with a very short impulse of duration of one second, and we normalize the result based on the incident bolometric luminosity:
\begin{equation}
\label{eq:resp}
    \psi_d(t,\lambda) = {1 \over {\Delta_t L_x}}\int_{S_{disk}} B_{\lambda}(T_{eff}(r(,t^{'}-\tau_{do}(x,y))ds.
\end{equation}
We do not subtract the flux from the non-irradiated disk, since we do not aim to linearize the equation and, in general, $\psi(t,\lambda)$ depends on the parameters, including $L_x$, which is a function of time.
When generating the response function, we used two types of time bins: linear and logarithmic -- as the linear time bin failed to capture the smallest delays created by the disk elements close to the black hole.  

To create the disk light curves from the incident X-ray light curves,  we generally do not use the concept of response function as defined by Equation~\ref{eq:resp}, but we calculate the result directly from Equation~\ref{eq:lum}.  In this approach, the amplitude of the irradiating flux can be arbitrarily large in comparison with the locally dissipated radiation flux.

\subsection{Second reprocessor}

\begin{figure}
\centering
\includegraphics[angle=270,width=0.5\textwidth ]{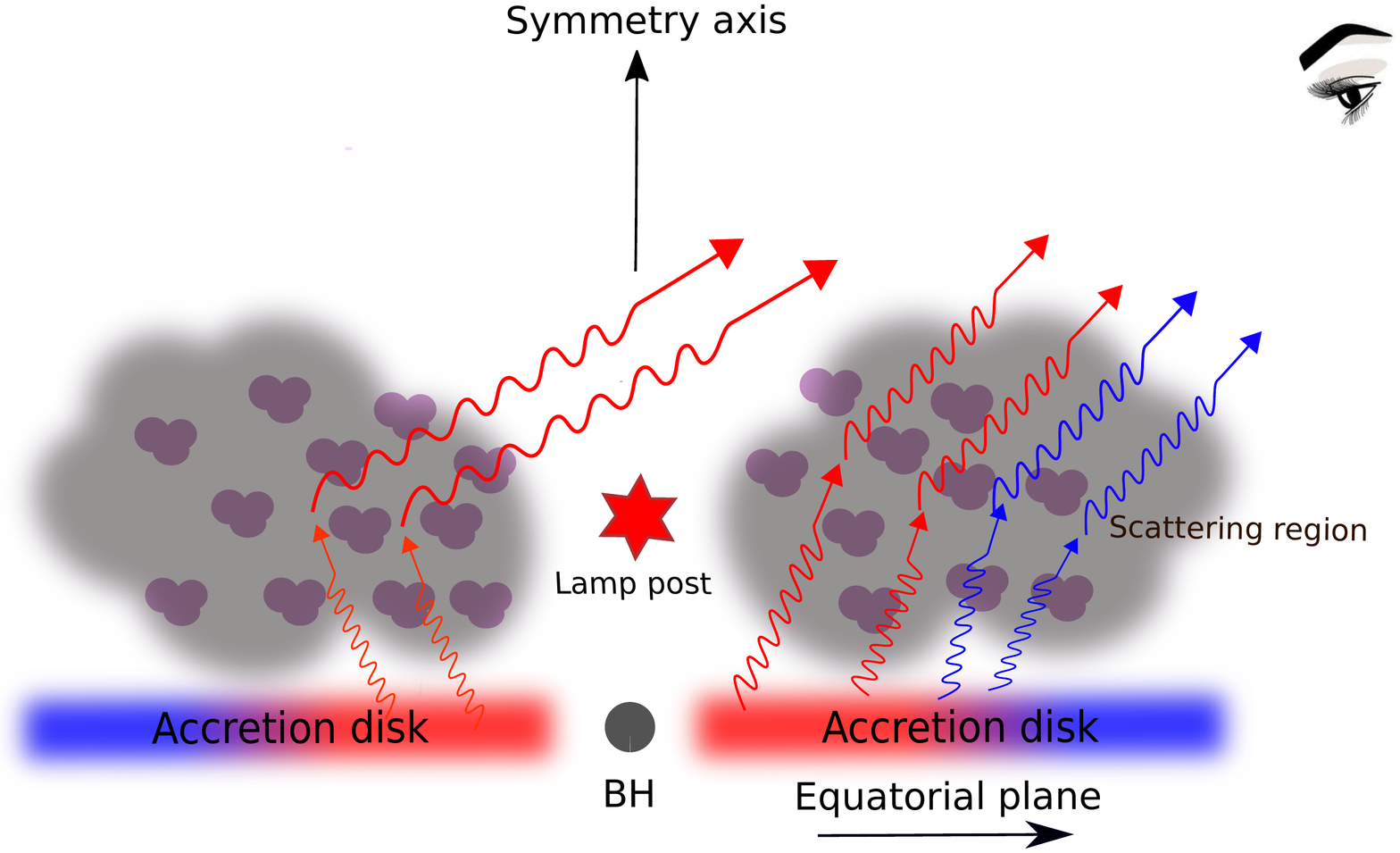}
\caption{\label{fig:geometry}Geometry of the reprocessing by the extended disk and extended BLR. We include only the scattering from the inter-cloud medium.}
\end{figure}

The BLR is a secondary medium responsible for the reprocessing of the irradiating flux \citep{lawther2018,korista2019,chelouche2019,netzer2021}. This medium is the source of the emission lines and the emission line time delays have been measured by many authors since many years using the reverberation mapping technique \citep[e.g.,][]{blandford_mckee82,peterson88,peterson93,  kaspi2000, peterson2004, bentz2013, dupu2014, grier17, martinez-aldama2019, du2019, panda_2019_frontiers, zajacek2021}. This process has mostly been modeled and measured as an independent process, neglecting the disk reprocessing. However, as pointed recently by many authors \citep{Korista_Goad_2001, korista2019,netzer2021, GuoH_etal_2022}, BLR is also the source of diffuse continuum, including Balmer continuum and scattering, which also vary -- this, together with some level of line contamination affects the measured continuum time delays. 

In this paper, we focus on the simplest aspect of the BLR, which is electron scattering of the photons by the inter-cloud BLR medium. Such a process does not imprint any characteristic features as a function of wavelength but can effectively modify the predicted net time delay between the two continuum bands. The schematic illustration of the geometry is shown in Figure~\ref{fig:geometry}. We include this effect through the Thomson scattering approximation. This means that the scattering does not change the photon frequency.

\begin{figure*}
\centering
\includegraphics[width=0.3\textwidth]{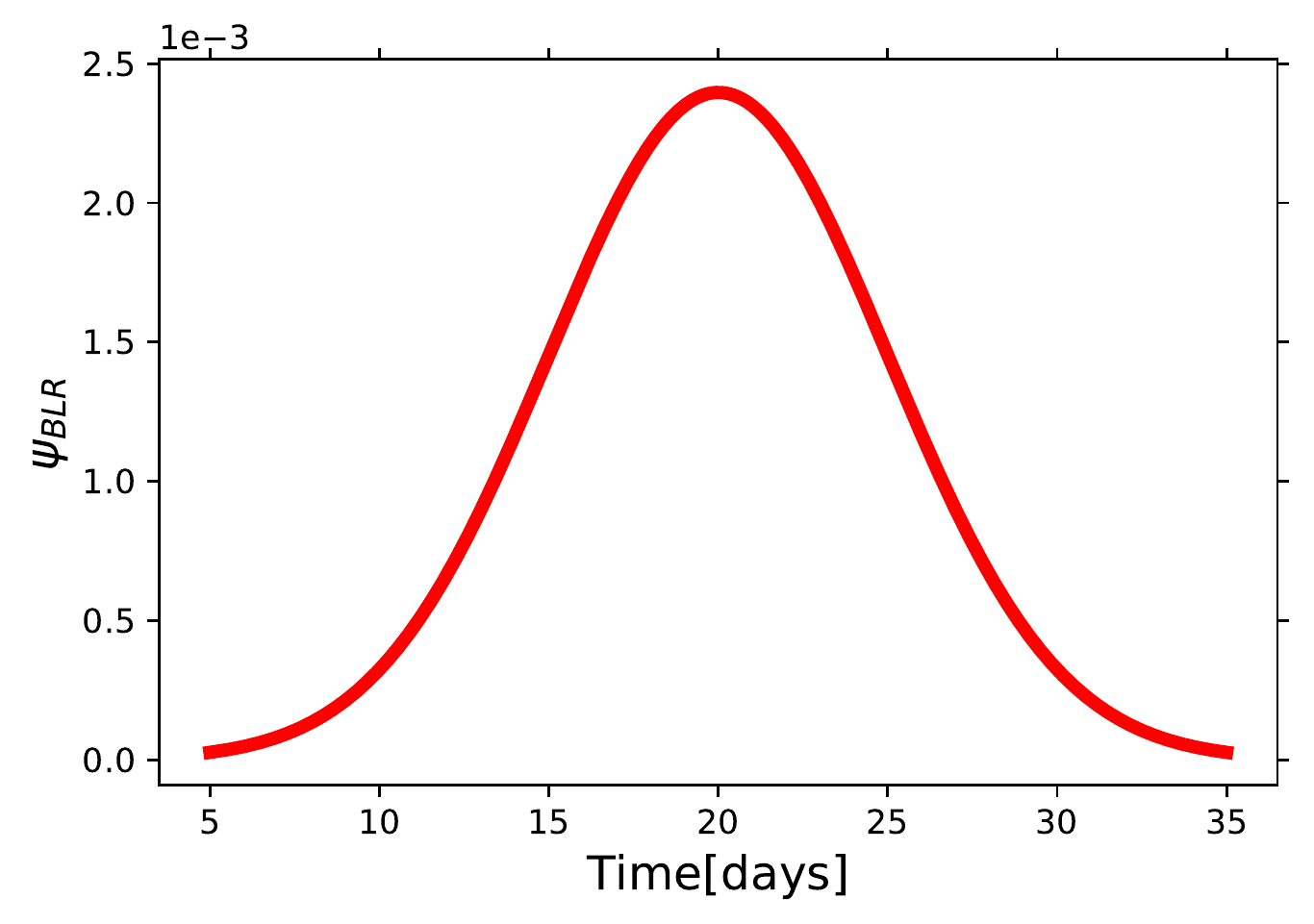}
\includegraphics[width=0.3\textwidth]{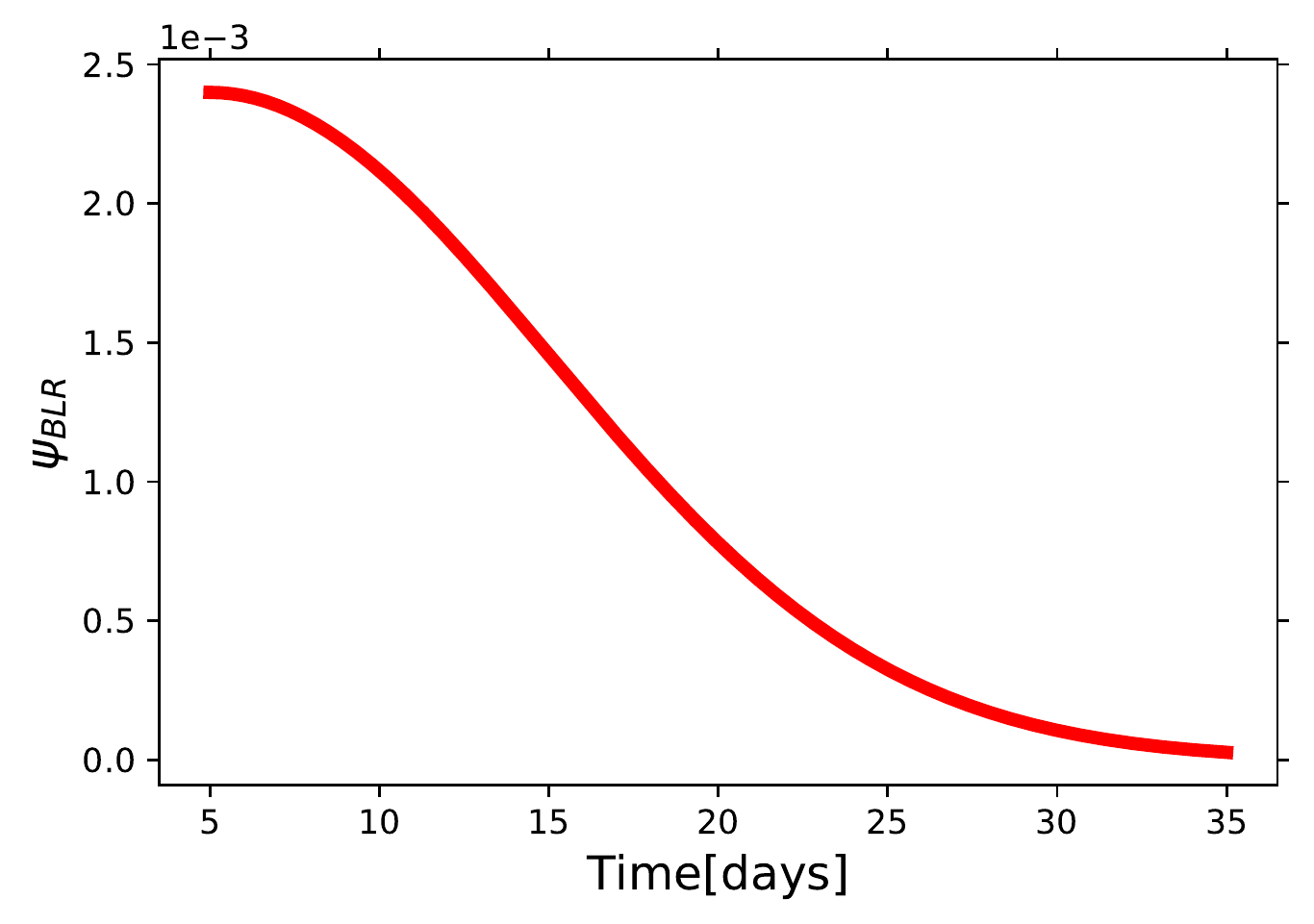}
\includegraphics[width=0.3\textwidth]{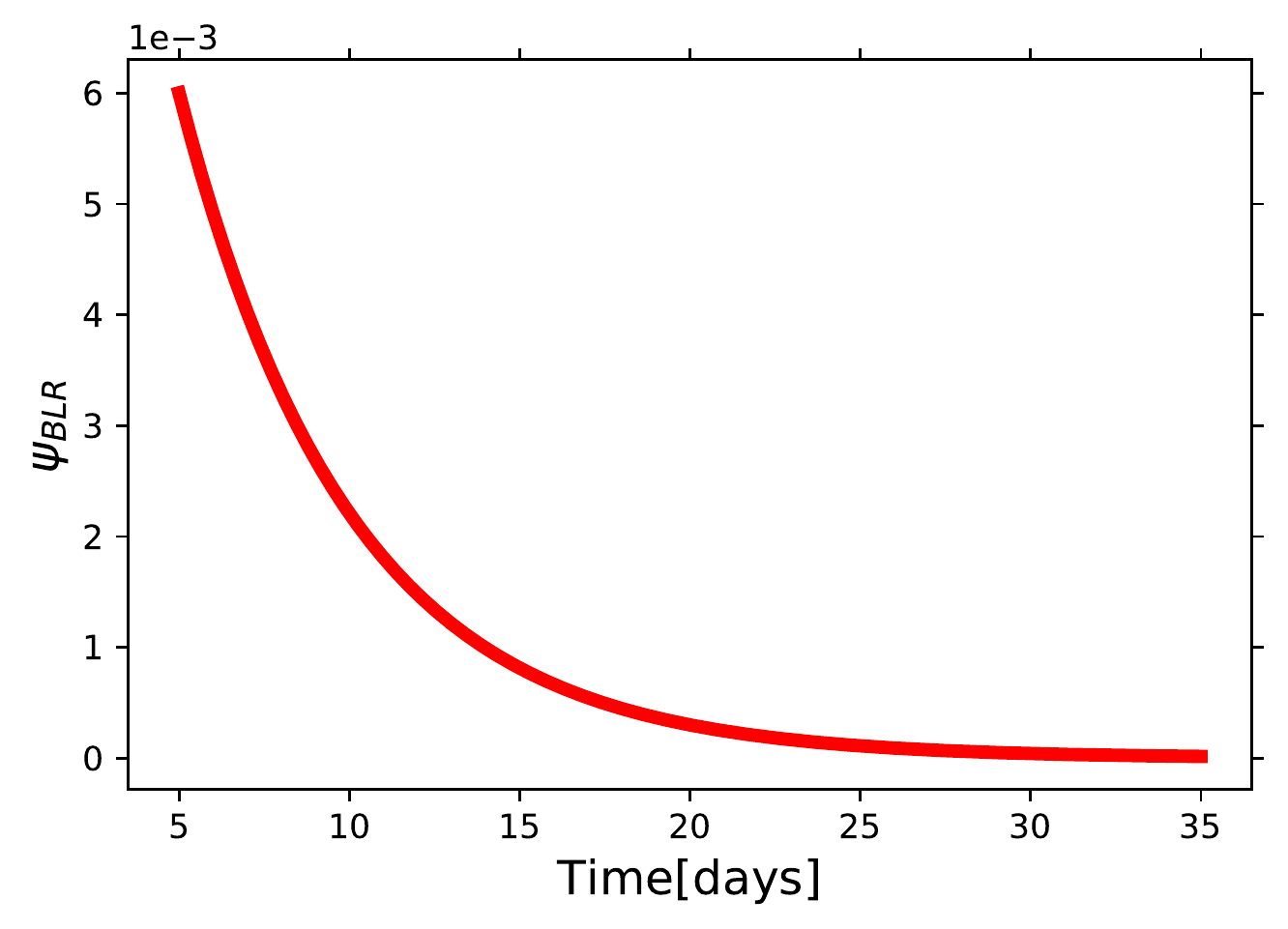}
\caption{\label{fig:3responses}BLR response function shapes used in this paper. Left panel: Full Gaussian ($\tau_{peak} = 20$ days, $\sigma = 5$ days); Middle panel: Half-Gaussian ($\tau_{peak} = 5$ days, $\sigma = 10$ days); Right panel: Exponential decay ($\tau_{peak} = 5$, decay rate$ = 0.1 $). In all three cases,  $t_0 = 5$ days, $t_{max} = 35$ days. }
\end{figure*}

In order to model the scattering effect, we used a simple analytical parametrization of the response of the second reprocessor, $\psi_{BLR}$. There are no simple direct observational determinations of such a response, but we can look for suitable parameters by looking at the measured responses for H$\beta$ lines \citep[e.g.,][]{grier2013,xiao2018,dupu2018} and assume that the inter-cloud medium follows a similar distribution. Thus, for the shape of this function, we assumed one symmetric function in the form of a Gaussian and two asymmetric functions: half-Gaussian and an exponential decay. The  peak location of the function in the time axis is given by $T_{peak}$. The non-zero contribution is included starting at a specific time delay $t_0$ up to $t_{max}$. The width of the Gaussian, $\sigma,$ or the decay timescale of the exponential ($t_{decay}$) are the parameters of the model. Exemplary shapes are presented in Figure~\ref{fig:3responses}. We assume the same shape of the $\psi_{BLR}$ for all the disk photons, independently from their wavelength and location on the accretion disk.

In the case of the presence of the second reprocessor, we introduce a parameter $f_{BLR}$ which weights the relative contribution from the two regions.  The parameter $f_{BLR}$ accounts for the fraction of photons scattered by the inter-cloud medium and introduces an additional time  delay. The factor (1 - $f_{BLR}$) represents the fraction of disk photons that reaches the observer without any scattering or additional time delay. In case of $f_{BLR}$ different from zero the total time delay is a combination of two processes: initial delay due to light travel time from the X-ray source to the disk and the extra time delay due to scattering in extended BLR inter-cloud medium.

 The assumption of extended BLR medium modeled $\psi_{BLR}$ implies that the final result is a convolution of these two effects:

\begin{equation}
\label{eq:convol}
\psi(\lambda,t) = (1 - f_{BLR}) \psi_d(\lambda,t) + f_{BLR} \int_{t_0}^{t_{max}} \psi_d(\lambda,t')\psi_{BLR}(t') dt'.
\end{equation}

The value of  $f_{BLR}$ can vary from 0 (no BLR reprocessing) to 1 (no disk reprocessing). The usually expected value is rather in the range from 0.1 to 0.3, if measured by an estimated solid angle of the BLR. However, fully ionized medium can have a different spatial distribution than the BLR denser clouds.

If we aim to calculate the effect of reprocessing from the long X-ray lightcurve, we apply the response function directly to our photon table \textbf{P(t',${\boldsymbol{\mathrm{\lambda}}}$)}, using the same parameter $f_{BLR}$: a fraction of $(1 - f_{BLR})$ remains unchanged and the fraction of $f_{BLR}$ is smeared by $\psi_{BLR}(t)$.

\subsection{Time delay measurements}

We considered two mathematical approaches to determine the expected time delay. In the first method, we use a single flare event (not a delta function but of final duration of 1 sec), we determined the response function by the disk for the corresponding parameters and we eventually added the response function of the second reprocessor. In this case  (see Section \ref{sect:single}), the whole light curve is not created, so this method is most accurate but does not adequately represent how the time delay is measured in the actual observational data. In the second approach, we created a realistically sampled light curve (incident and in each of the bands), and the time delay is measured using methods comparing the two light curves (see Section \ref{sect:long}).  Independently from the method, all delays are always measured with respect to the X-ray flare event.

\subsubsection{Single-event delay}
\label{sect:single}

In the case of a single event, we constructed the standard response function $\psi$ for the accretion disk with the lamp post geometry (see Section~\ref{sect:disk}), combined it with properly normalized response function from the second reprocessor, and calculated the expected time delay using the formula \citep{koratkar1991}:
\begin{equation}
\label{eq:responce}
\tau(\lambda) = \frac{\int t\psi(t,\lambda)dt}{\int 
\psi(t,\lambda)dt}
.\end{equation}

Computations in this case require much denser time grids in comparison with the computations for long X-ray light curves (see  Section~\ref{sect:disk}), since the onset of the reprocessed flare has to be well resolved in this case and that happens very close to the black hole.  

\subsubsection{Realistic light curves}
\label{sect:long}

In this case, we simulated the entire incident radiation curve, with realistic sampling, and determined the observed continuum curves at selected wavelengths.  Three methods are most frequently adopted: interpolated cross-correlation function, ICCF \citep{peterson1993,Sun_etal_2018}, JAVELIN \citep{Zu_etal_2011, Zu_etal_2013, Zu_etal_2016}, and $\chi^2$ \citep{Czerny_etal_2013,Bao2022}. In the present paper, we concentrate on the first one (ICCF), which brings rather stable results.

\subsection{Mkn 110 as a motivation for the adopted parameters in simulations}

Complex variability was recently discussed in detail for the source Mkn 110 \citep{mkn1102021}. So, in order to put our simulations onto realistic footing, we predominantly focus on parameters well representing this source and the light curve duration and spacing characteristic for this source. Mkn 110 \citep{Mkn1969} is a well-studied nearby optically bright, radio-intermediate (R$\approx$1.6) narrow-line Seyfert 1 galaxy (NLS1s) at a redshift z = 0.036 \citep[see e.g.,][]{dasgupta2006}. As implied by the relativistically broadened X-ray emission line (O {\sc vii}), the cold, standard disk extends there at least up to 20-100 $r_g$ \citep{reeves2021} and Fe K$\alpha$ line study implied that the cold disk down to 1.24 $r_g$ is needed \citep{mantovani2016}, although we note that such a fit was achieved for unlikely inclination of 80 deg. A more recent study of combined data from XMM-Newton and NuSTAR gave more conservative values of $\sim 20 r_g$ \citep{porquet2021}. The level of polarization in the optical band in this source is low \citep[$\sim$0.5\%,][]{afanasiev2019}, but the low polarization does not necessarily imply the low optical depth of the scattering medium \citep[e.g.,][]{Sniegowska_etal_2022} and the presence of very highly ionized medium is revealed through the presence of Fe {\sc xxvi} emission line \citep{mantovani2016}.  The viewing angle is estimated at $18.0 \pm 3.1$ deg. \citep{afanasiev2019}. The black hole mass in this source was estimated to be $2 \times 10^7 M_{\odot}$ \citep{Bentz_Katz_2015} and also adopted by \citet{Vincentelli2021}. Older measurements have claimed a higher value, $1.2 \times 10^8 M_{\odot}$, and even larger values has been determined from the recent polarization method, $\log M = 8.32 \pm 0.21 M_{\odot}$ \citep{afanasiev2019}.
 The source has been monitored with Neil Gehrels Swift Observatory \citep{2004ApJ...611.1005G}, with a cadence roughly once per day, for about 200 days, and we adopted this setup in our simulations. A detailed disk reverberation is studied by \citet{mkn1102021}, using good cadence data from Neil Gehrels Swift Observatory and LCO network and they found the variability on two different time scales in Mkn 110. The variability time scale below ten days is mostly consistent with accretion disk reverberation with a maximum two-day lag between the shortest wavelength (W2 band) and longest $z$ band. On the longer time scale they found that the $g$ band lags the hard X-ray (from BAT) by ten days and a similar lag was also observed between the $z$ and $g$ band which is not consistent with the disk reverberation. The author proposed that the longer time scale and higher time delay can be due to the contamination from the diffuse emission of the BLR. \citet{vincentelli2022} also discussed the effect of X-ray luminosity on the lag spectra. During the low X-ray luminosity state, they do not observe any u-band excess and negative time lag excess as frequently seen in many AGNs. However, at high a X-ray luminosity state, the u-band excess is visible, which is dedicated to the diffuse BLR emission. However, these authors also argue that the excess lag in X-ray can also be explained by moving the corona to farther distances. In the same sense, our study also sees the possibility to disentangle the BLR and corona height contribution in order to explain the disk reverberation.  
\section{Results}

We presents the result of our project aimed at testing the time delay in the presence of the two reprocessors using simulated light curves. This allows us to test under which circumstances, if at all, the time delays from the disk alone can be recovered. 

\subsection{Response function of the disk and single event delay}
\label{sect:resp_from_1}

\begin{figure}
\centering
\includegraphics[width=0.5\textwidth]{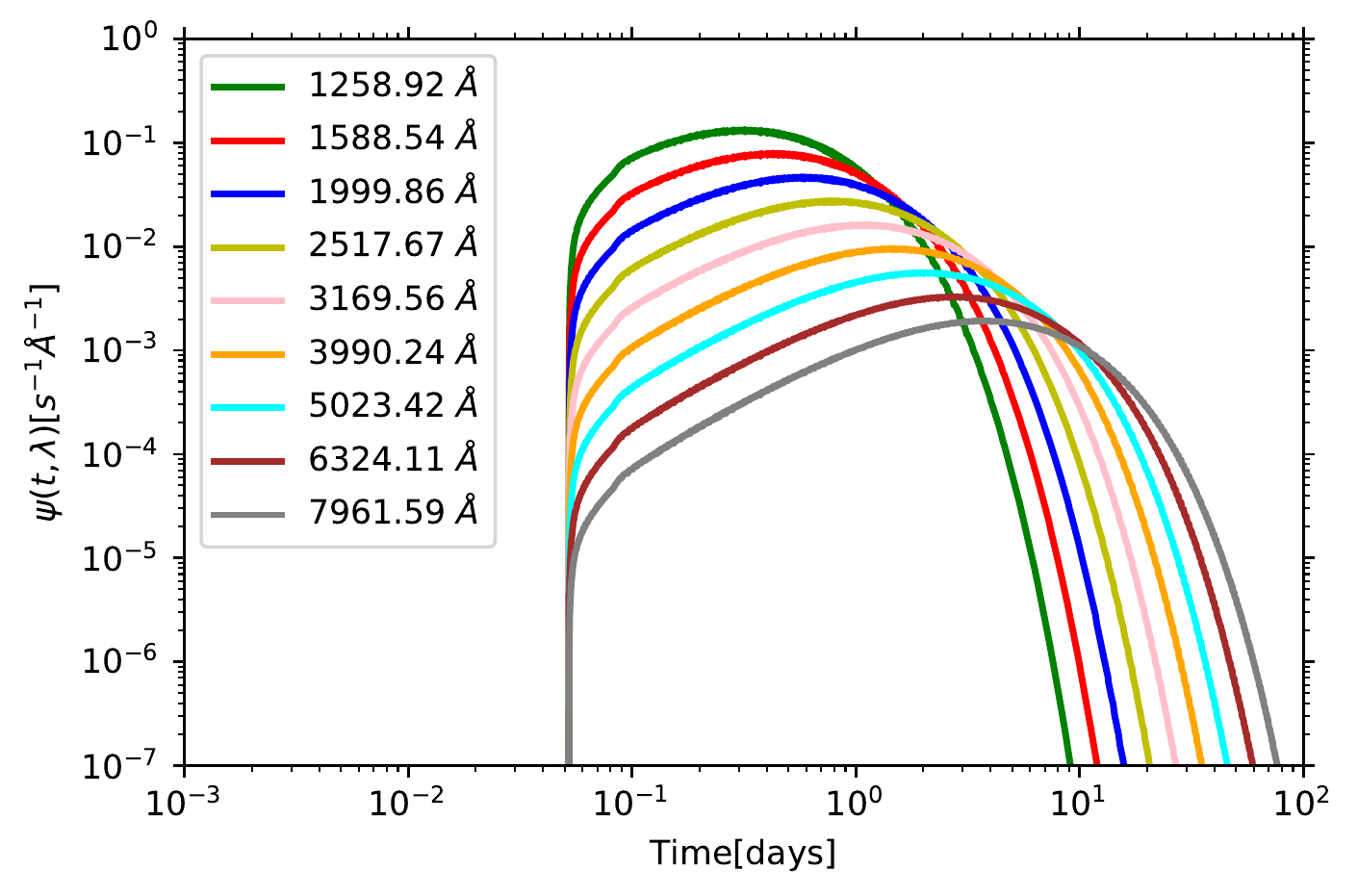}
\caption{\label{fig:response_disk}Response function shapes for the accretion disk at selected wavelengths (see text). Parameters: black hole mass = $10^{8}M_{\odot}$, Eddington ratio = 1.0, $L_X = 10^{40}$ erg s$^{-1}$, height: $h = 5 r_g$, and viewing angle: $i = 30$ degrees.}
\end{figure}

\begin{figure}
\centering
\includegraphics[width=0.5\textwidth]{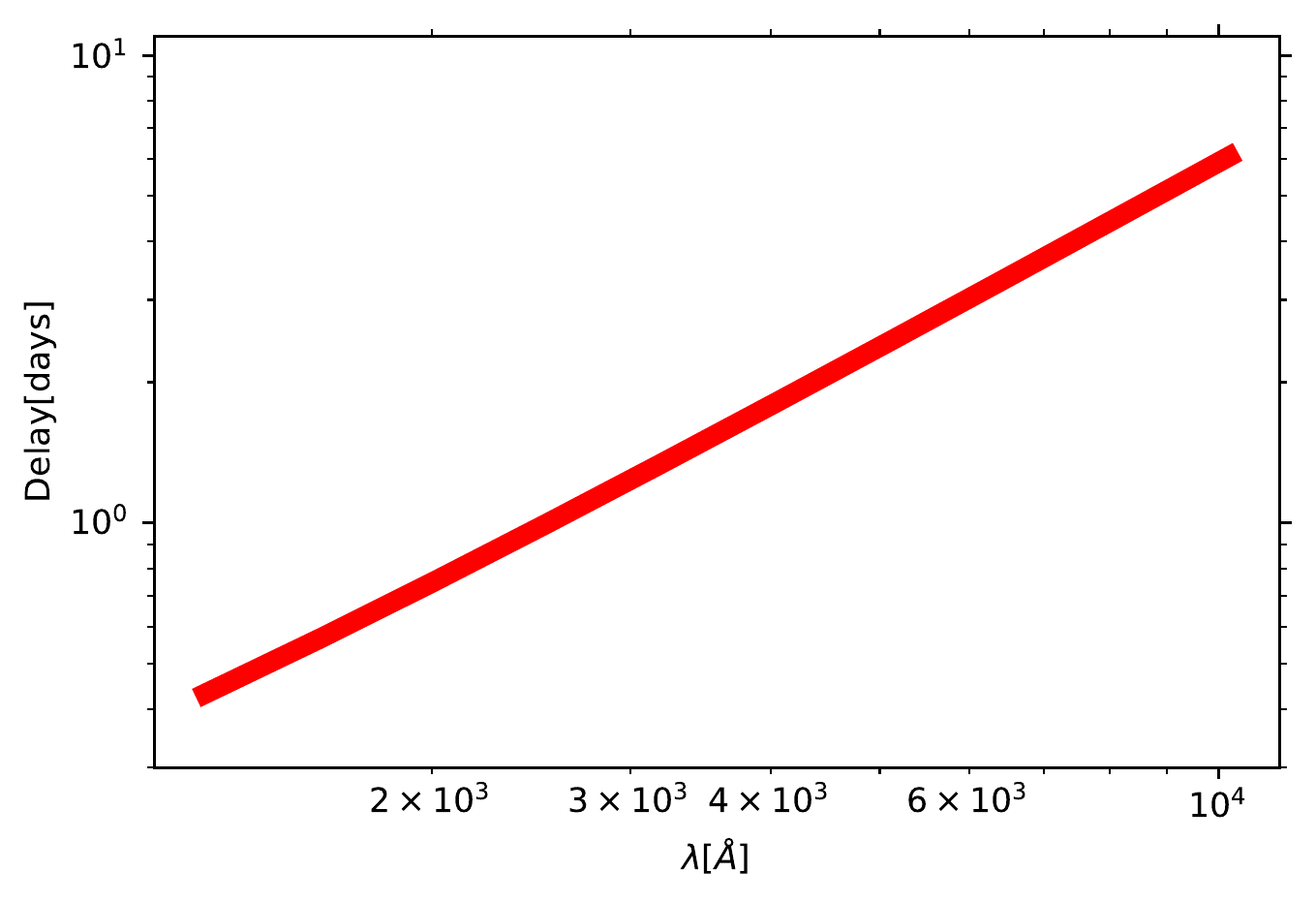}
\caption{\label{fig:delay1} Time delays calculated from the disk response functions from Fig.~\ref{fig:response_disk}.  Parameters as in Fig.~\ref{fig:response_disk}.}
\end{figure}

We first calculated one example of the response function, that is, the result of the reprocessing of the delta signal from the X-ray lamp by the disk, without any presence from the second reprocessor. Such computations require much denser grids in space and time as specified in Sect.~\ref{sect:disk} to adequately see the onset of the radiation. For this exercise, we used the accretion disk model with the following parameters: $M = 10^8 M_{\odot}$, $L/L_{Edd} = 1$, lamp height of $H = 5 R_g$, the lamp luminosity of $L_X = 10^{40}$ erg s$^{-1}$, and the viewing angle of $i = 30$ deg. 
All the delays are calculated with respect to the corona. Although we can generate a response function for any wavelength, we usually store and show the response function only for nine values of $\lambda$, from 1258.92 $\AA$ (response1) to 7961.59 $\AA$ (response9), adopting a constant logarithmic step.
We show the results in Fig.~\ref{fig:response_disk}. The overall shape is similar to the responses derived by \citet{kammoun2021}, although we do not have  GR corrections. We comment  more quantitatively on this issue in Appendix~\ref{appen}. 

We calculated the centroid times of all response functions using Equation~\ref{eq:responce}. 
The results (see Figure~\ref{fig:delay1}) are consistent with the simple analytic formula of \citet{collier1999}. We also compared the normalization of the best fit $\tau \propto \lambda^{4/3}$ trend with the expectations from \citet{collier1999}. Their formula contains an unspecified factor $X$ which accounts for the peak contribution to the total emission from a given radius through a scaling of $X = h c /(k T \lambda)$, which they estimated to be of the order of 3-4. Our value, derived from numerical computations, gives $X = 2.47$, much lower than the factor suggested by \citet{collier1999}, but higher than the semi-analytical relation ($X = 1.65$) proposed by \citet{siemiginowska1989}.

\subsection{Response function from two reprocessors and single event delay}
\label{sect:two_repro}

We go on to calculate the time delay from Equation~\ref{eq:responce} as a function of the wavelength, for a combined disk and BLR effect and for several values of the parameter $f_{BLR}$.

\begin{figure}
\centering
\includegraphics[width=0.5\textwidth]{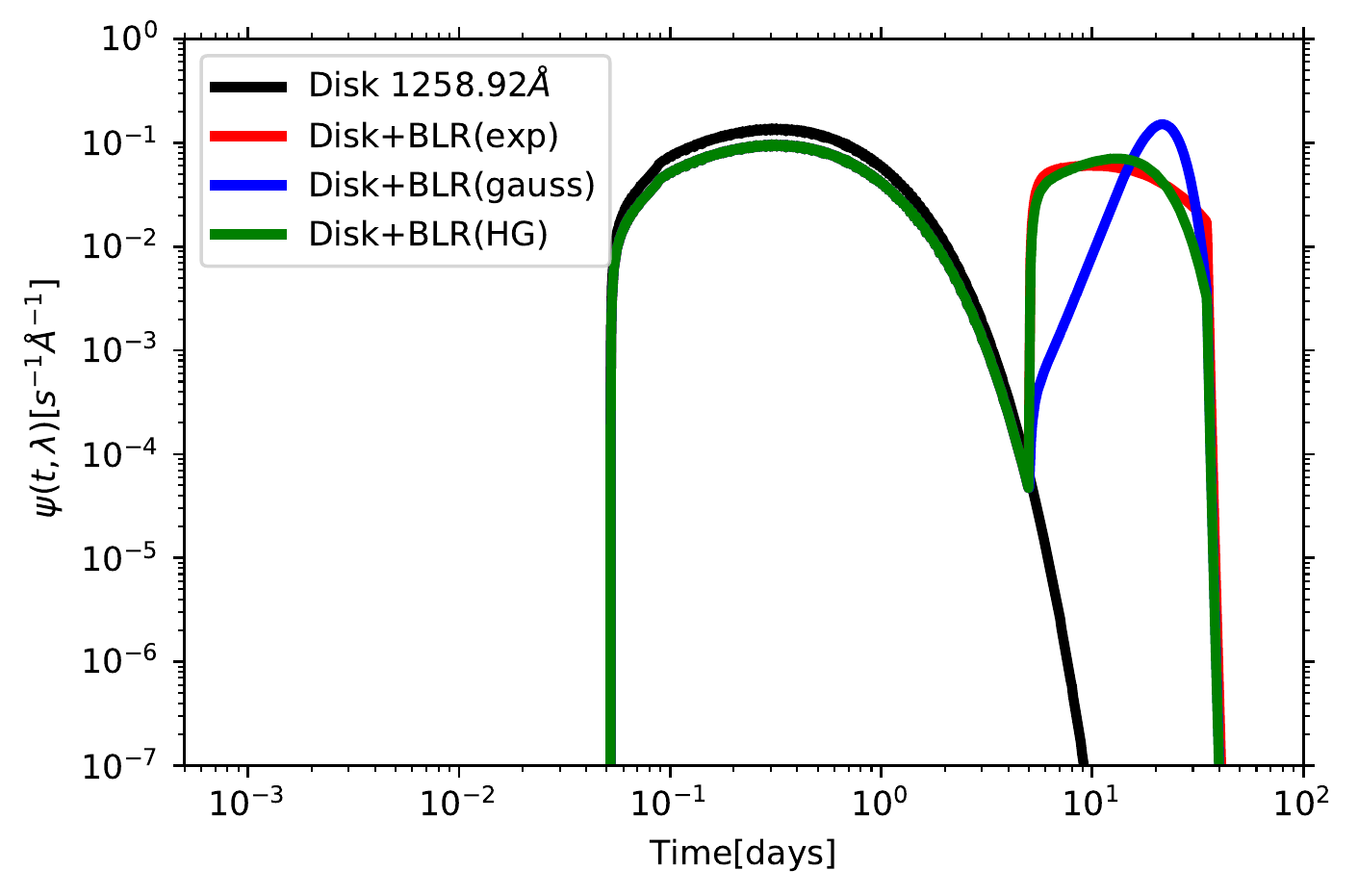}

\caption{\label{fig:response_disk_blr}Comparison between disk response function and disk plus BLR response function for different BLR profiles (Gaussian, half-Gaussian, and exponential; see Fig.~\ref{fig:3responses}), and 30\% BLR contribution is used. Parameters: black hole mass $10^{8}M_{\odot}$,  Eddington ratio = 1.0, $L_X = 10^{40}$ erg s$^{-1}$ , height $h = 5 r_g$, viewing angle $i = 30$ degrees).}
\end{figure}

\begin{figure}
\centering
\includegraphics[width=0.5\textwidth]{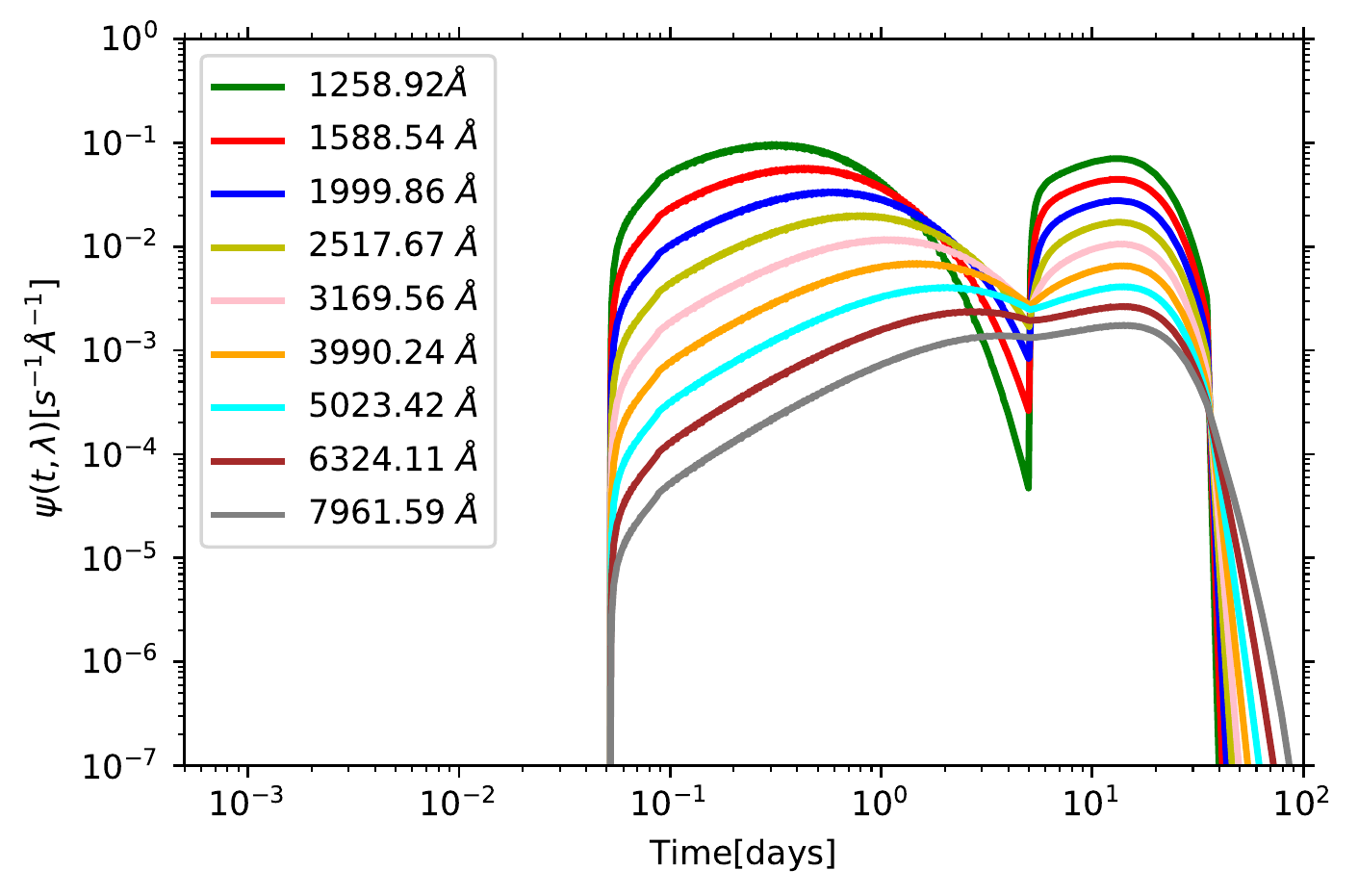}

\caption{\label{fig:response_all_blr}Combined response function, disk plus BLR, for BLR profile half-Gaussian (see middle panel in Fig.~\ref{fig:3responses}) at selected wavelengths. Parameters: black hole mass $10^{8}M_{\odot}$, Eddington ratio = 1.0, $L_X = 10^{40}$ erg s$^{-1}$ , height $h = 5 r_g$, viewing angle $i = 30$ degree, $f_{BLR} = 30\%$.}
\end{figure}
\begin{figure}
\centering
\includegraphics[width=0.5\textwidth]{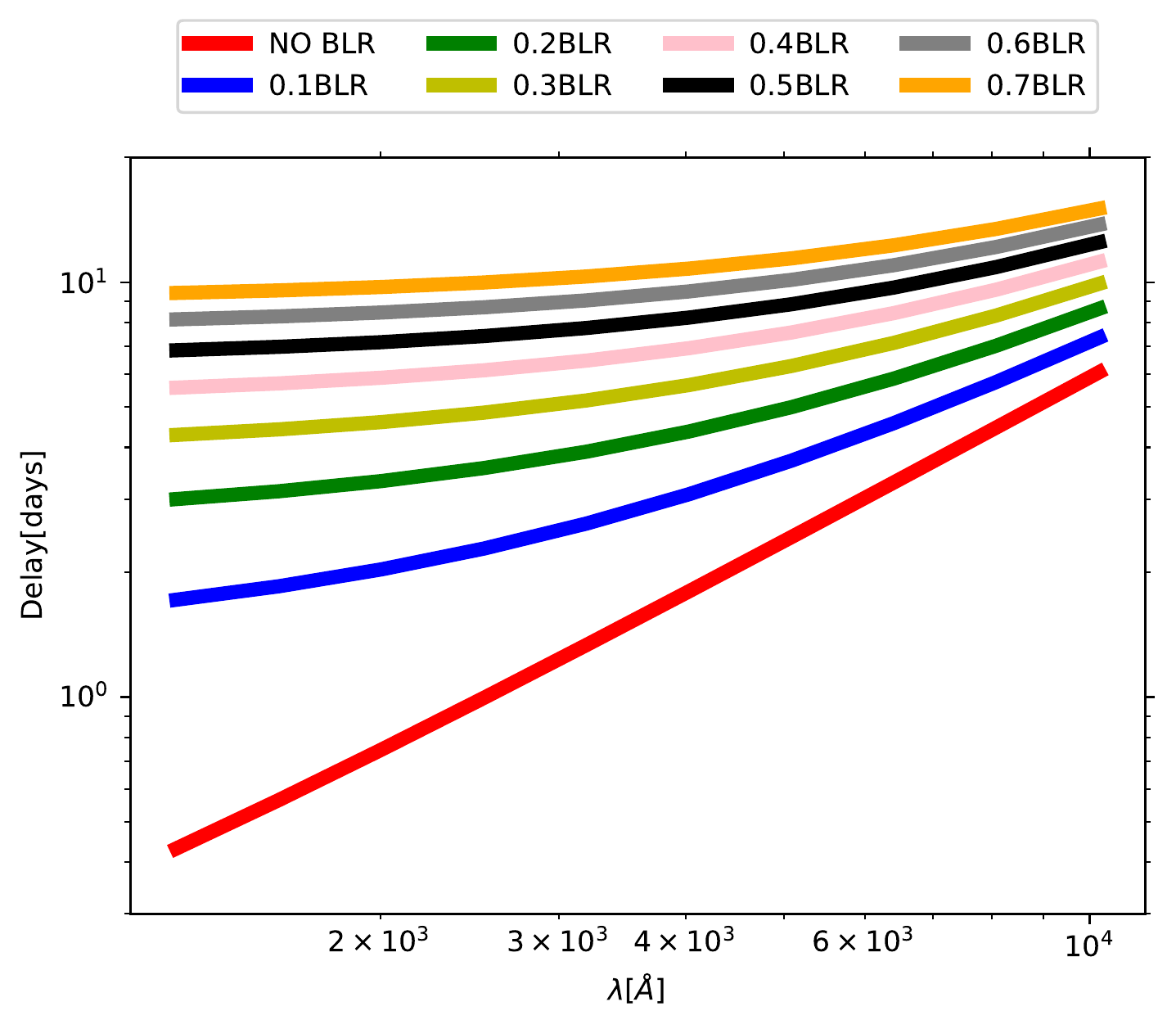}
\includegraphics[width=0.5\textwidth]{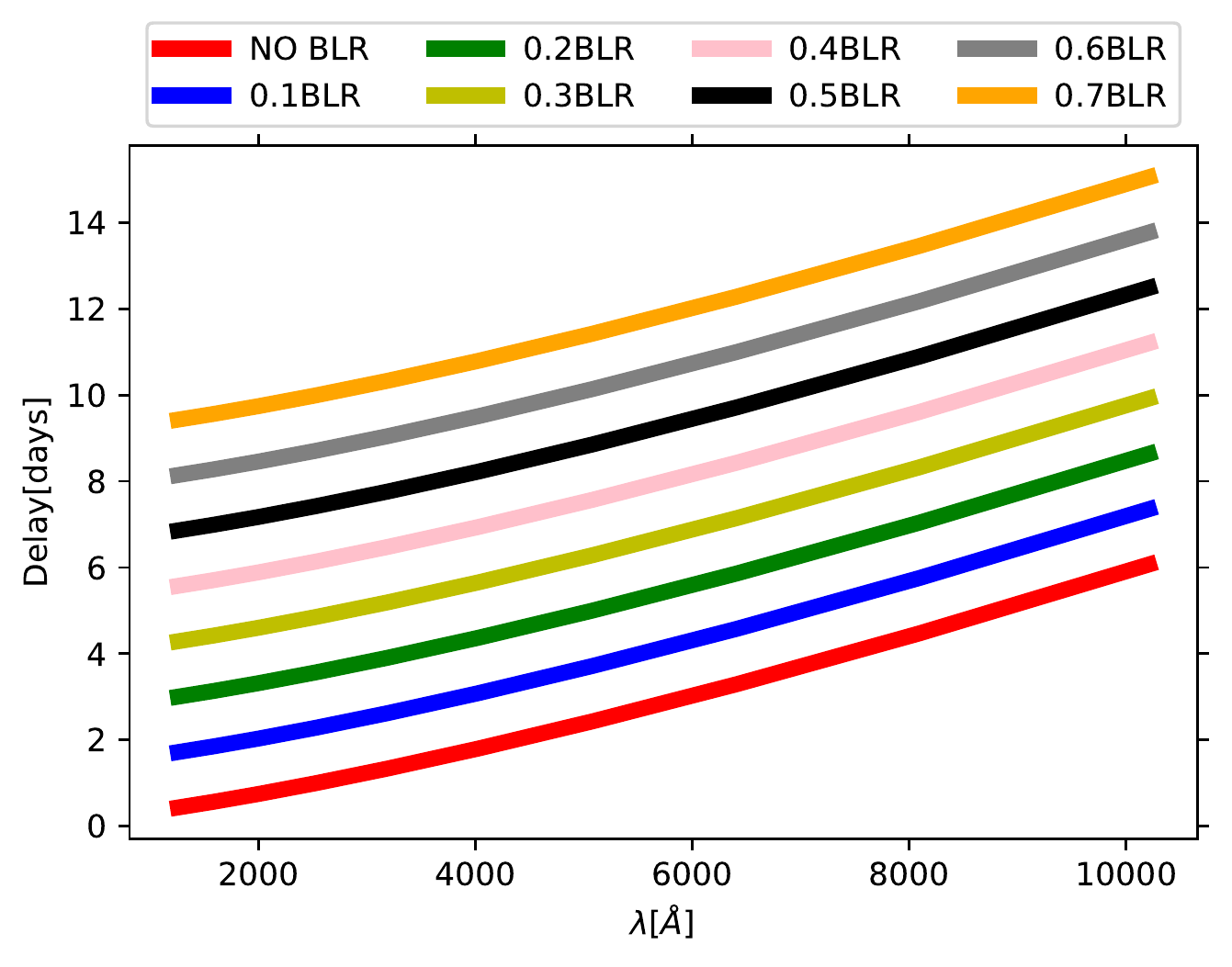}
\caption{\label{fig:all_lags}Delay for all response function with different BLR percentage contribution for BLR profile half-Gaussian (see middle panel in Fig.~\ref{fig:3responses}). Upper panel: log scale, lower panel: linear scale. Parameters: black hole mass $10^{8}M_{\odot}$, Eddington ratio = 1.0, $L_X = 10^{40}$ erg s$^{-1}$ , height $h = 5 r_g$, viewing angle $i = 30$ degree.}
\end{figure}

The exemplary response functions for the two reprocessor setup is shown in Figure~\ref{fig:response_disk_blr}. We plot the shape for the shortest wavelength only, but for three shapes of the BLR response illustrated in Figure~\ref{fig:3responses}. 

We see that in the two-reprocessor setup, the combined response has two peaks and the shape of the new response depend on the adopted description of the BLR. In Figure~\ref{fig:response_all_blr}, we plot the response functions for representative wavelengths, selecting half-Gaussian (Figure~\ref{fig:3responses}, middle panel) that represents the BLR. We see that the deep valley between the two peaks becomes more shallow when we go towards longer wavelengths and, finally, the two-peak structure disappears.  We note that the dependence of the time delay on wavelengths at the longest time delay is not due to wavelength-dependent effect in BLR itself (as we assume Thomson scattering in the inter-cloud medium), but it is due to the fact that those are photons generated at large disk radii, with the delay generated between the X-ray source and their origin and the net effect is a convolution, as given by Equation~\ref{eq:convol}.  

Next, for the same parameters, we calculated the time delays from two reprocessors. The results are shown in Figure~\ref{fig:all_lags}. For $f_{BLR} = 0$
the usual $\tau \propto \lambda^{4/3}$ is recovered. Now we add an offset in the Equation~\ref{eq:convol}, which corresponds to the fraction of BLR. The time delay for different contribution of BLR is estimated and in the linear scale the delays at higher BLR contribution are just shifted with respect to f$_{BLR}$ = 0 (lower panel of Figure~\ref{fig:all_lags}). However, the result appears to be very interesting when we plot the time delays in log-log space (upper panel of Figure~\ref{fig:all_lags}). With increasing $f_{BLR}$, we obtain more shallower relations than the standard one. 
So, not only does the time delay become  longer overall due to the extra scattering in the BLR region, but also the slope of the relation changes. In the extreme case, when the disk contribution becomes small and most of the photons are actually scattered, the delay only weakly depends on the wavelength. However, some dependence on the wavelength would remain even for 100\% BLR contribution since the photons (before going on to the observer) are first reprocessed in the disk with a wavelength-dependent delay. It is important to note that the change of the slope is actually seen only if we use log-log plot. 

\begin{figure}
\centering
\includegraphics[width=0.5\textwidth]{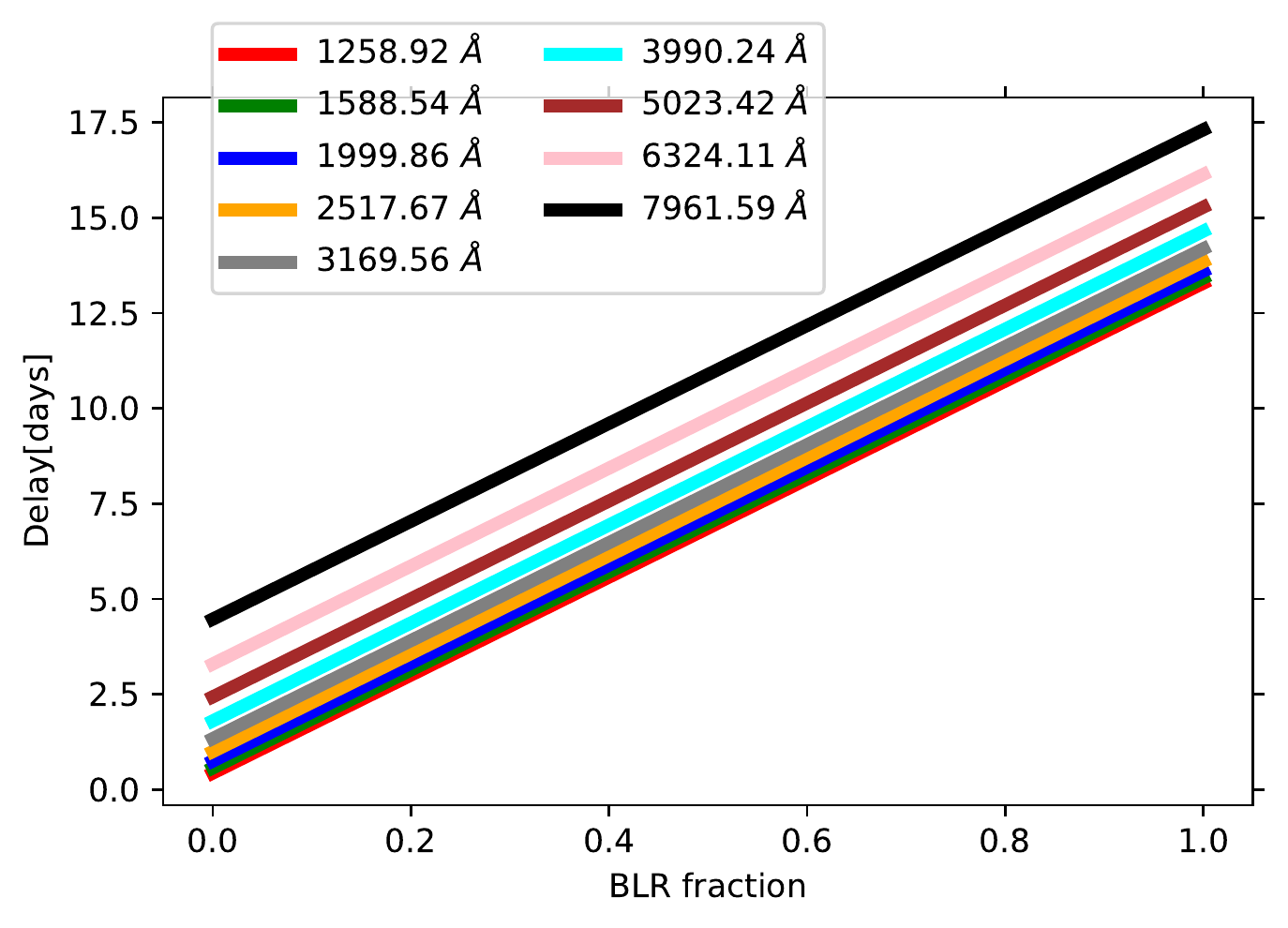}
\caption{\label{fig:netzer2}Time delay as a function of the BLR contribution for exemplary wavelengths. For BLR profile, we used half Gaussian as shown in the middle panel Fig.~\ref{fig:3responses}. Model parameters are as in Figure~\ref{fig:all_lags}.}
\end{figure}

It is also interesting to show the dependence of the time delay at selected wavelengths, but as a function of BLR contribution.  
\citet{netzer2021} argued (partially following \citealt{lawther2018}) that the delay from two reprocessors is a linear combination of the two time delays (disk and BLR) weighted with the flux contribution. We plot the expectations from our model in Figure~\ref{fig:netzer2}. 
The plot supports the claim of the linear dependence of the time delay on the BLR contribution for all wavelengths if this contribution comes from scattering. The dependence is indeed linear, with the time delay plot shifted up with the increase in the wavelength.

The contribution from the scattering by the inter-cloud medium in BLR can thus easily account for too large disk sizes claimed from the data. In addition, the change of the delay shape shown in Figure~\ref{fig:all_lags} might, in principle, reveal this effect in the data. However, \citet{kammoun2021_data} were able to fit well the observational data for a number of sources at the expense of postulating large height of the illuminating source.  Thus, our more general model -- which includes the arbitrary lamppost height and arbitrary contamination by the disk photon scattering -- might be degenerate with respect to these two parameters and, in the data fitting in the future, we will not be able to discriminate  among them. In order to address this problem in advance, in our simulations, we calculated the effect of these two parameters for a range of lamppost source luminosities. 

\begin{figure}
\centering
\includegraphics[width=0.5\textwidth]{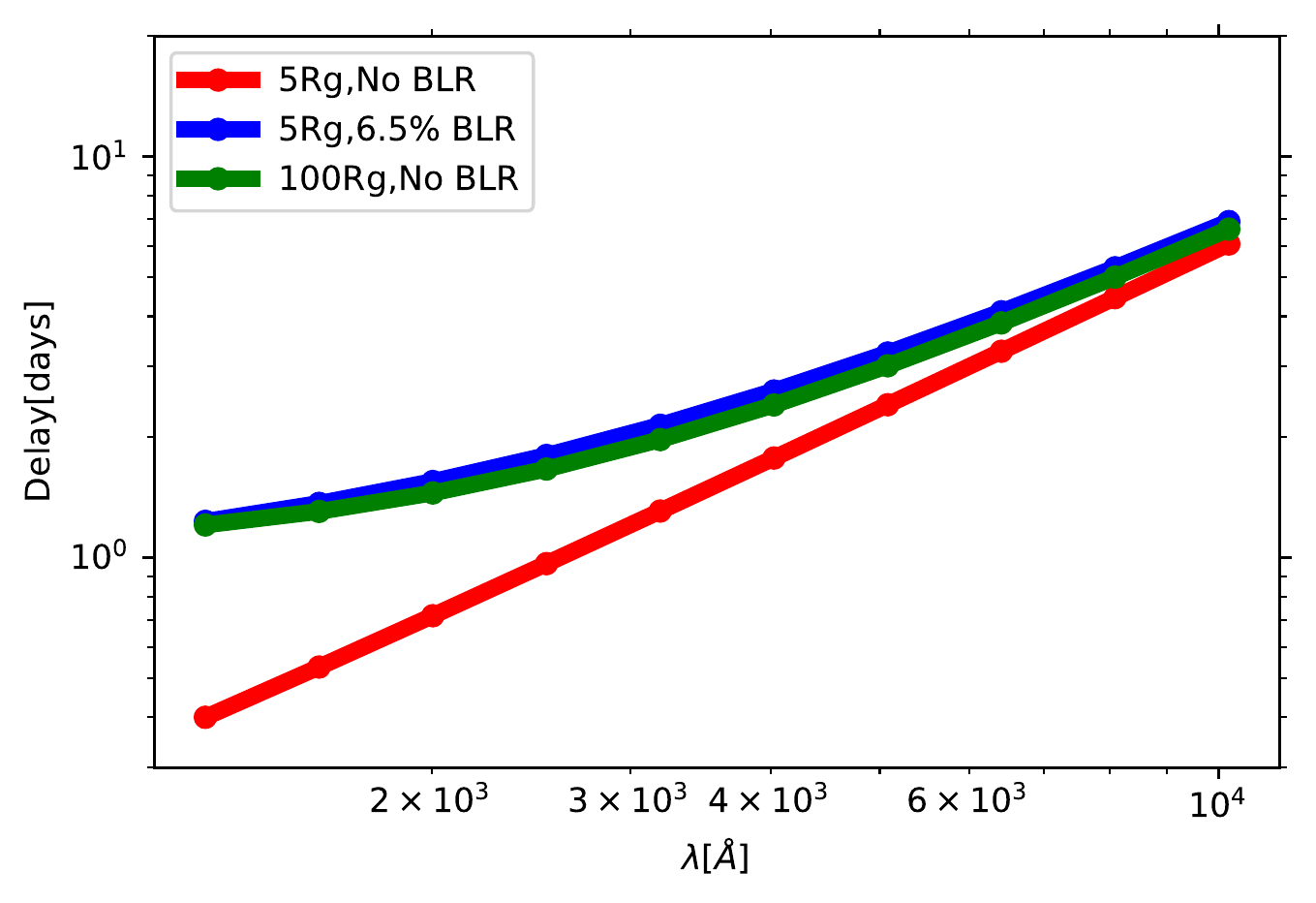}
\includegraphics[width=0.5\textwidth]{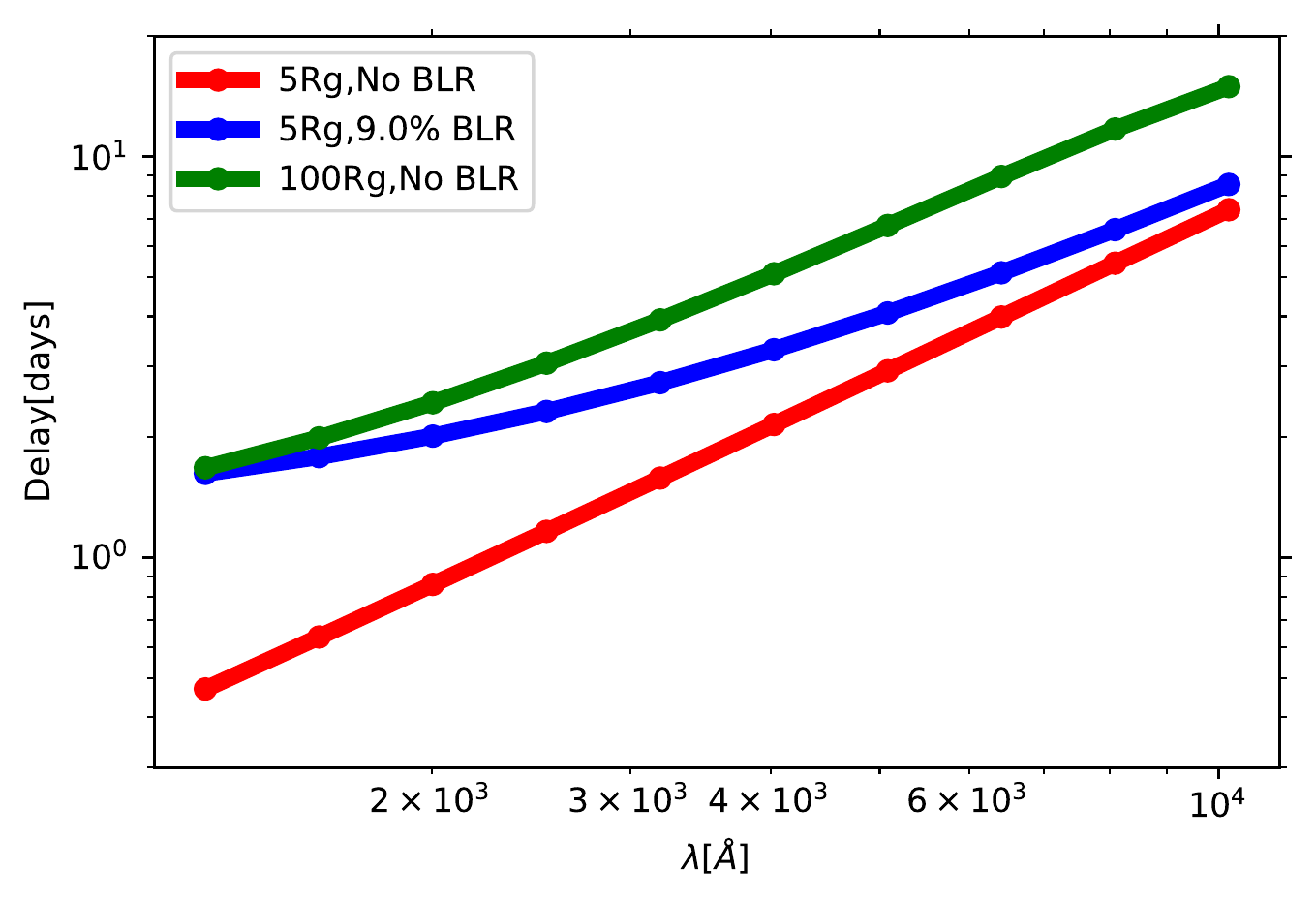}
\caption{\label{fig:H_BLR}Time delay as a function of the wavelength for models with different lamppost height and for no BLR and for BLR contribution. For the BLR profile, we used a half-Gaussian. as shown in the middle panel of Fig.~\ref{fig:3responses}. Model parameters (top): Black hole mass $10^8$ M$_{\odot}$, $L/L_{\rm Edd}$ = 1, inclination angle 30 degree, and luminosity of the corona is  $10^{40}$ erg $s^{-1}$. Model parameters (bottom): Black hole mass $10^8$ M$_{\odot}$, $L/L_{\rm Edd}$ = 1, inclination angle 30 degree, and luminosity of the corona is 3.78 $\times 10^{46}$ erg $s^{-1}$.}
\end{figure}

 First, we assumed a very faint lamppost of $10^{40}$ erg $s^{-1}$, and  we repeated the disk delay computations for several height values and compared the results with the expected time delay for small height but with BLR contribution. We see that a change of the lamppost height leads to a flattening of the delay curve, as shown by \citet{kammoun2021}. In particular, Fig. 18 of their paper shows that the effect is very similar to the introduction of the BLR scattering.

The curvature introduced to the delay curves in low-luminosity case is actually very similar in the case of an increased height or some BLR scattering (see Figure~\ref{fig:H_BLR}, upper panel). It is a serious source of the degeneracy in the future data fitting, although the relative importance of the two effects depends strongly on the parameters. For example, from our standard model ($\sigma = 10$ days, $L_X = 10^{40}$ erg s$^{-1}$, the contribution of BLR just $\sim 7 $\% gives the same effect as moving the lamp height from 5 to 100 $R_g$, and smaller values of $\sigma$ further reduce this factor. 

 Next, we repeated the simulations for the incident X-ray flux to 30\% of the disk bolometric luminosity. Such high luminosity ($L_X \sim 3 \times 10^{46}$ erg $^{-1}$) brings different result: 9\% BLR contribution delay is only matched at the smallest wavelength, not at longer wavelengths. Thus, the overall curvature becomes  different (see Figure~\ref{fig:H_BLR}, lower panel), which opens up the possibility to differentiate the effect of the height from the effect of the BLR scattering in the data. This shows that for X-ray-bright sources, we can differentiate between the lamppost height and the BLR contamination if the data is of sufficient quality, but it might be much more difficult for X-ray weak sources.

\subsection{Time delays from light curves}

In the case of observational data, we do not have a direct insight into the response function; however, some techniques allow us to recover it from the data. The observed light curves depend not only on the system parameters, but also on the sampling, while the measured time delay depends not only on the light curves but on the method to determine the time delay. Therefore, we repeated our experiment from  Sections~\ref{sect:resp_from_1} and \ref{sect:two_repro}, using the long artificial light curves with different adopted power and data sampling. As a standard, we adopted the frequency break timescale of 75 days, sampling of one day, and the ICCF method. Computations are based on ten statistical realizations of the process, allowing us to show the errors representing the dispersion, namely, a likely error in a single measurement.

\subsubsection{Role of the frequency break in the power spectrum in disk time delay measurement}

\begin{figure}
\centering
\includegraphics[width=0.5\textwidth]{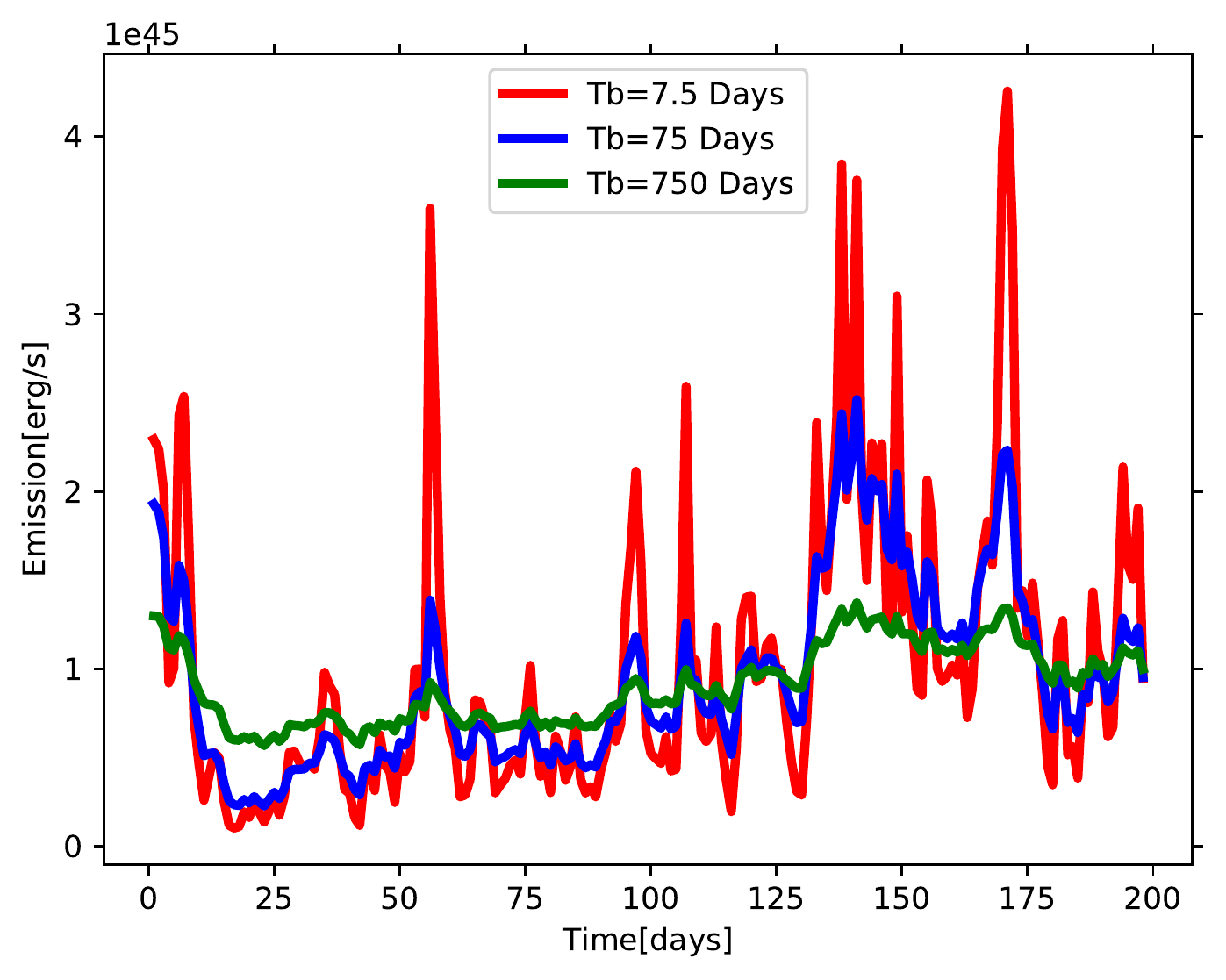}
\caption{\label{fig:X-ray-curves}Different X-ray light curve generated for different breaking time. Red, blue, and green are for breaking times of 7.5 days, 75 days, and 750 days, respectively.}
\end{figure}

First, again for test purposes, we calculated the examples of X-ray light curves representing different intrinsic timescales assumed in parametrization of the power spectrum. We adopted the time step of one day, duration of 200 days, and the irradiating X-ray lightcurve was calculated as described in Section~\ref{sect:curves}. We assumed the level of X-ray variability of set by normalized dispersion of 0.3 in the whole light curve of duration of $10^8$\;s. The examples of the light curve for three values of the high frequency break are shown in Figure~\ref{fig:X-ray-curves}. We see that a small value for the timescale corresponding to the frequency break gives much sharper values of the curve peaks and much higher variability amplitude in a period of 200 days (i.e., much shorter than the whole curve duration). All three curves were obtained from the same value of the parameter initializing random generator -- for a better comparison.

\begin{figure}
\centering
\includegraphics[width=0.43\textwidth]{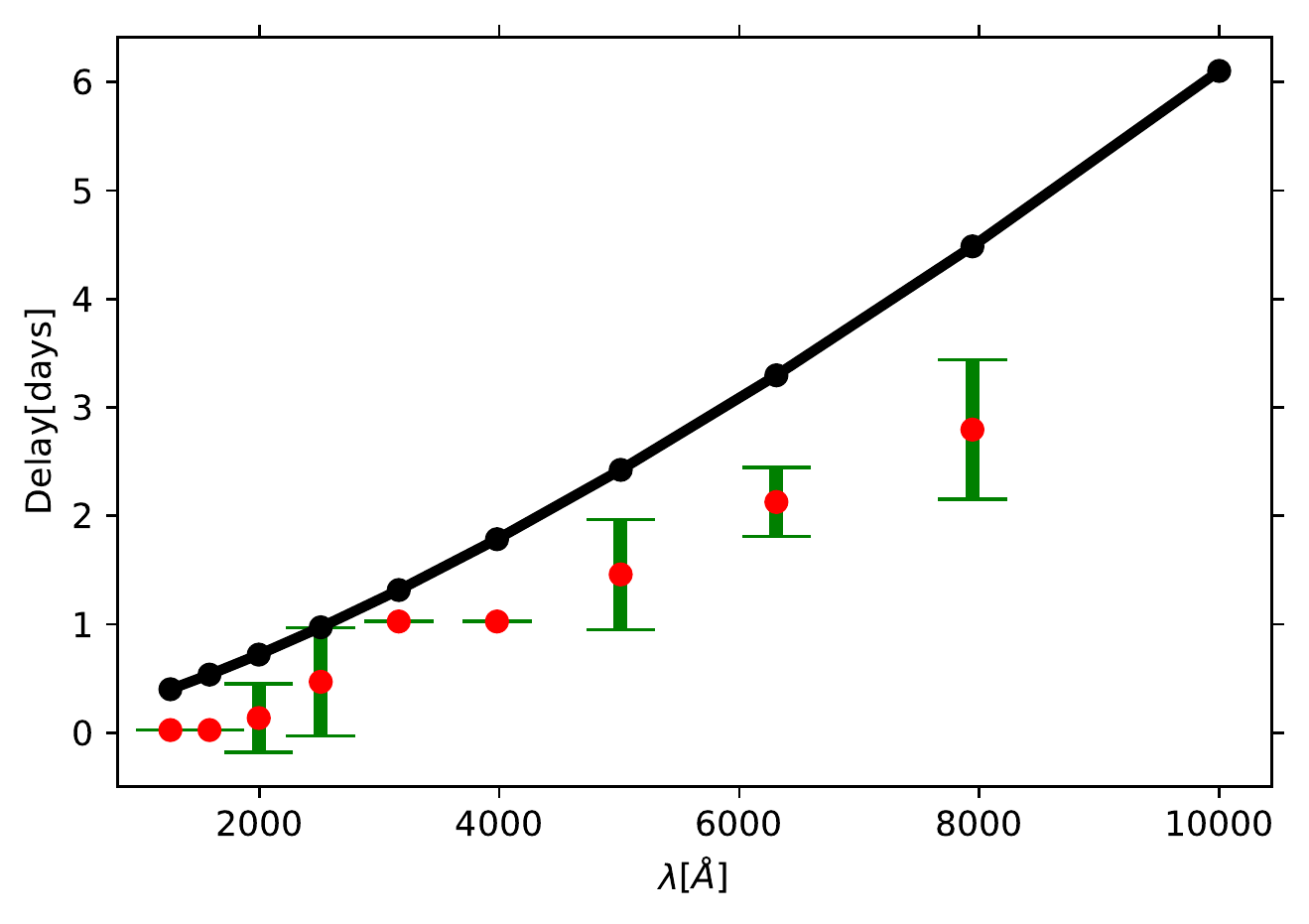}
\includegraphics[width=0.43\textwidth]{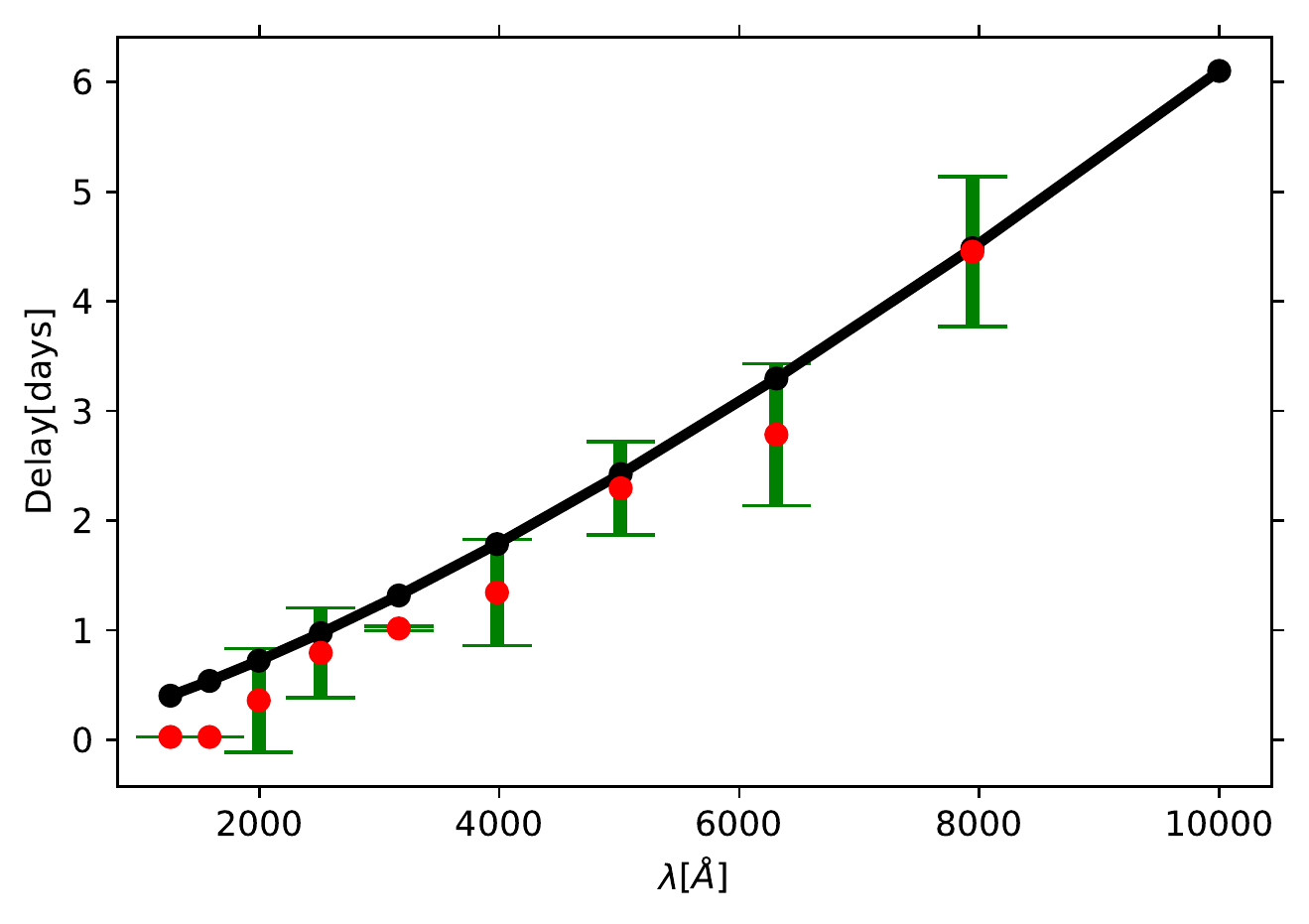}
\includegraphics[width=0.43\textwidth]{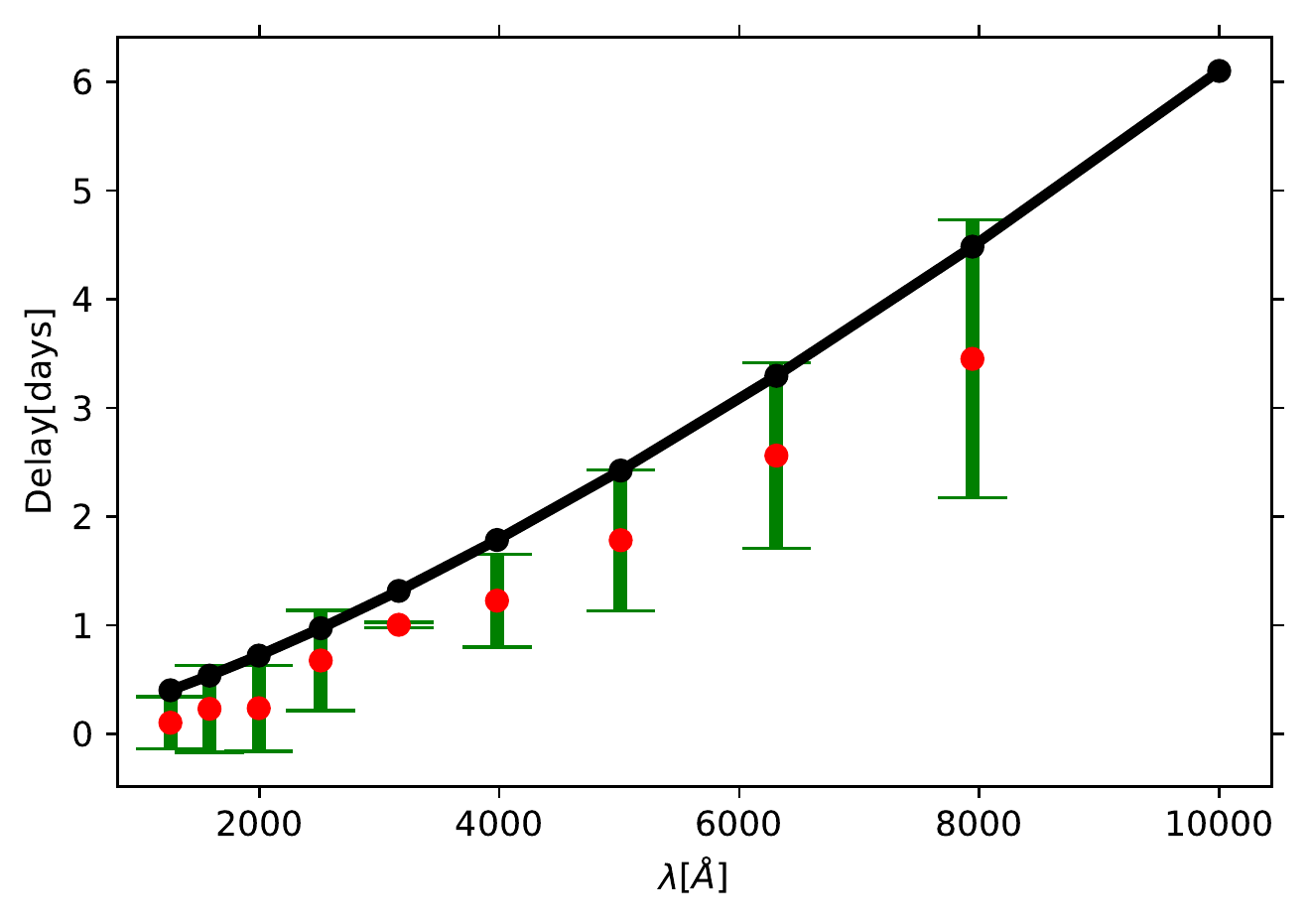}
\caption{\label{fig:delay_disk_curve} Time delay as a function of the wavelength calculated with ICCF method from the artificial light curves for the frequency break corresponding to timescales of 7.5 day, (upper panel), 75 days (second panel), and 750 days (third panel). Model parameters: Black hole mass $10^8 M_{\odot}$, Eddington ratio = 1.0, corona height = 5 r$_g$, inclination angle = 30 degree, and luminosity is on the order of 10$^{45}$ erg s$^{-1}$. Black points and continuous line represents the delays expected from the disk response function.
}
\end{figure}

 The X-ray curve seems more smooth when the timescale corresponding to high frequency break is longer, but otherwise, the geometry of the system is not affected. In order to check whether this indeed could affect the measured time delay, we calculated the delay for the three values of the frequency breaks, using ICCF.  The results are shown in Figure~\ref{fig:delay_disk_curve}. 
 
 Comparing the time delays obtained from the light curves to the delays calculated directly from the response function (Figure~\ref{fig:delay1}, upper panel), we see that for a black hole mass, $10^8 M_{\odot}$, the time delays are relatively  well recovered  
only in the case of a 75-day characteristic time variability (middle panel of Figure~\ref{fig:delay_disk_curve}). When variability is faster, the numerically calculated time delays are systematically much too short in comparison with expectations. If the variability timescales are longer, the numerical delays are marginally consistent with expectations within an error, but they again locate themselves systematically below the expected values. At the shortest wavelengths, even the optimum variability timescale underestimates the delay, but this is directly caused by the adopted one-day sampling which is not enough to resolve the innermost part of the disk. Increasing the incident X-ray flux to $\sim 3 \times 10^{46}$ erg s$^{-1}$ does not improve the results. 
We discuss the issue later.

This trend to obtain numerically the time delays which are shorter than expected is rather interesting and potentially important for actual data analysis. The optimum characteristic variability of 75 days is within the duration of the total lightcurve of $\sim 150$ days, so characteristic peaks are a few and apparently well sampled. If the actual timescale is much longer than the total observing time, we may have no strong features to rely on for time delay measurement. Indeed, if we use the light curves of the duration of 1000 days (again, with a one-day sampling) and the remaining parameters unchanged, the agreement between the numerical results and predictions is much better. Additionally, the asymmetry in the time reprocessing by the disk can also contribute if there are only very few strong peaks in both curves. On the other hand, if the characteristic timescale is much shorter than the total observing run and the sampling rate is not very dense the curve is too noisy.
It might thus be recommended to check the characteristic timescale in the data (e.g., using the structure function) and compare it to the derived time delay in order to additionally discuss  the potential bias in the time lag determination.

\subsubsection{Bin size effect in the disk time delay measurements}

\begin{figure}
\centering
\includegraphics[width=0.5\textwidth]{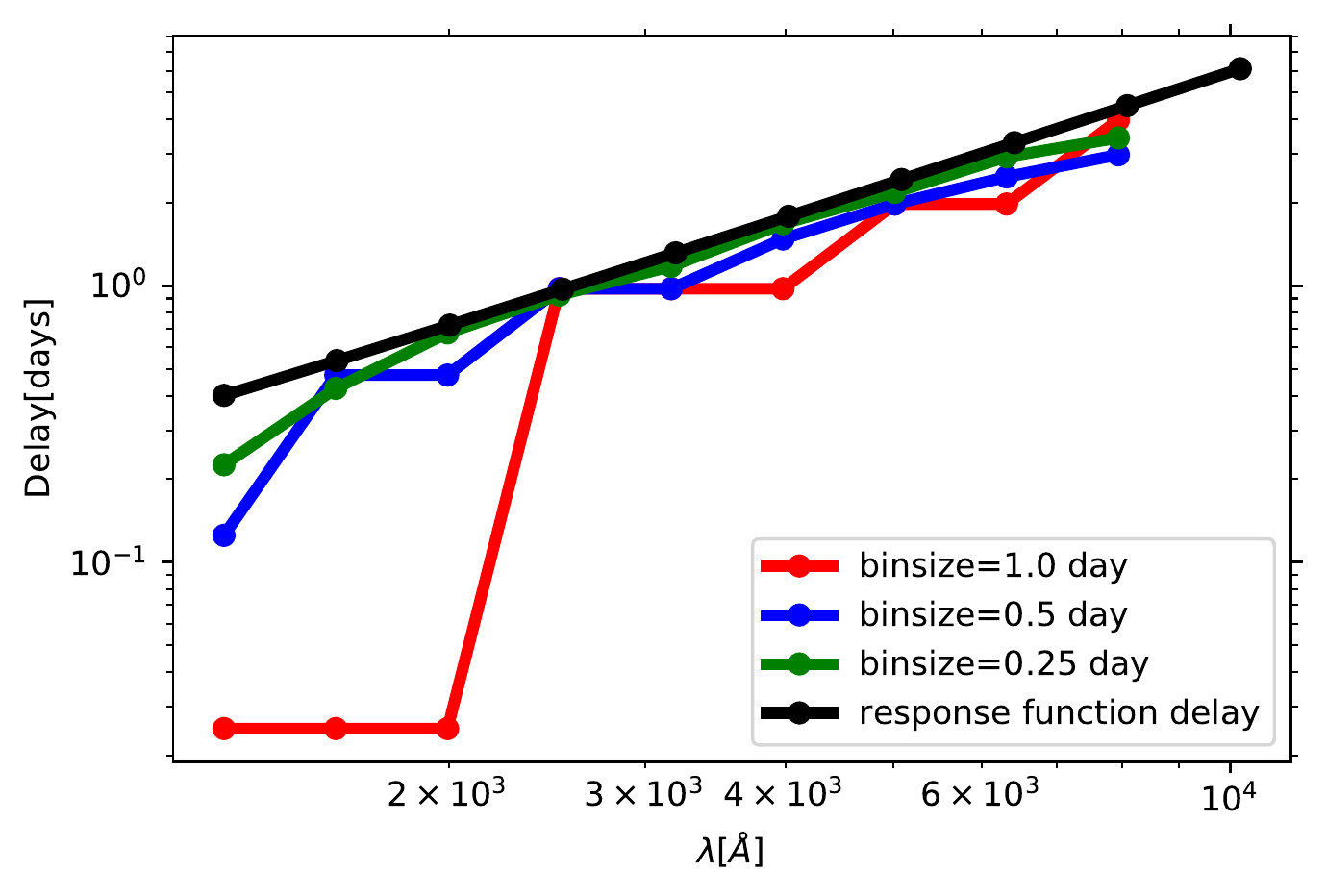}
\caption{\label{fig:delay_bin_size}  Time delay as a function of the wavelength calculated with ICCF method from the artificial light curves for different bin size. Black curve is expected delay from the response function. Model parameters: Black hole mass $10^8 M_{\odot}$, Eddington ratio = 1.0, corona height = 5r$_g$, inclination angle = 30 degrees, and luminosity is on the order of 10$^{45}$ erg s$^{-1}$.}
\end{figure}

 As we show in Figure~\ref{fig:delay_disk_curve}, the prediction of the delays from the simulated light curves systematically underestimate the delay by $\sim 30$\% at 4000 \AA, and the effect is stronger at the shortest wavelengths.  

 In order to see whether this is the result of inadequate sampling, we repeated the analysis for just one frequency break corresponding to 75 days, but for an increased data sampling and keeping the total length of the curves unchanged -- effectively increasing the number of observational points. The effect is shown in Figure~\ref{fig:delay_bin_size}. Indeed, with the denser sampling the numerical light curve time delay was systematically approaching the expected response of the disk. Already, the sampling of 0.5 day was enough to measure well the time delay at 2000 \AA\  and longer wavelengths, for the adopted black hole mass of $10^8 M_{\odot}$. The shorter wavelengths $\sim 1000$ \AA~ would require still denser sampling, as even 0.25 of a day would still underestimate the delay almost by a factor of 2.

\subsection{Time sampling and the black hole mass}

\begin{figure}
\centering
\includegraphics[width=0.5\textwidth]{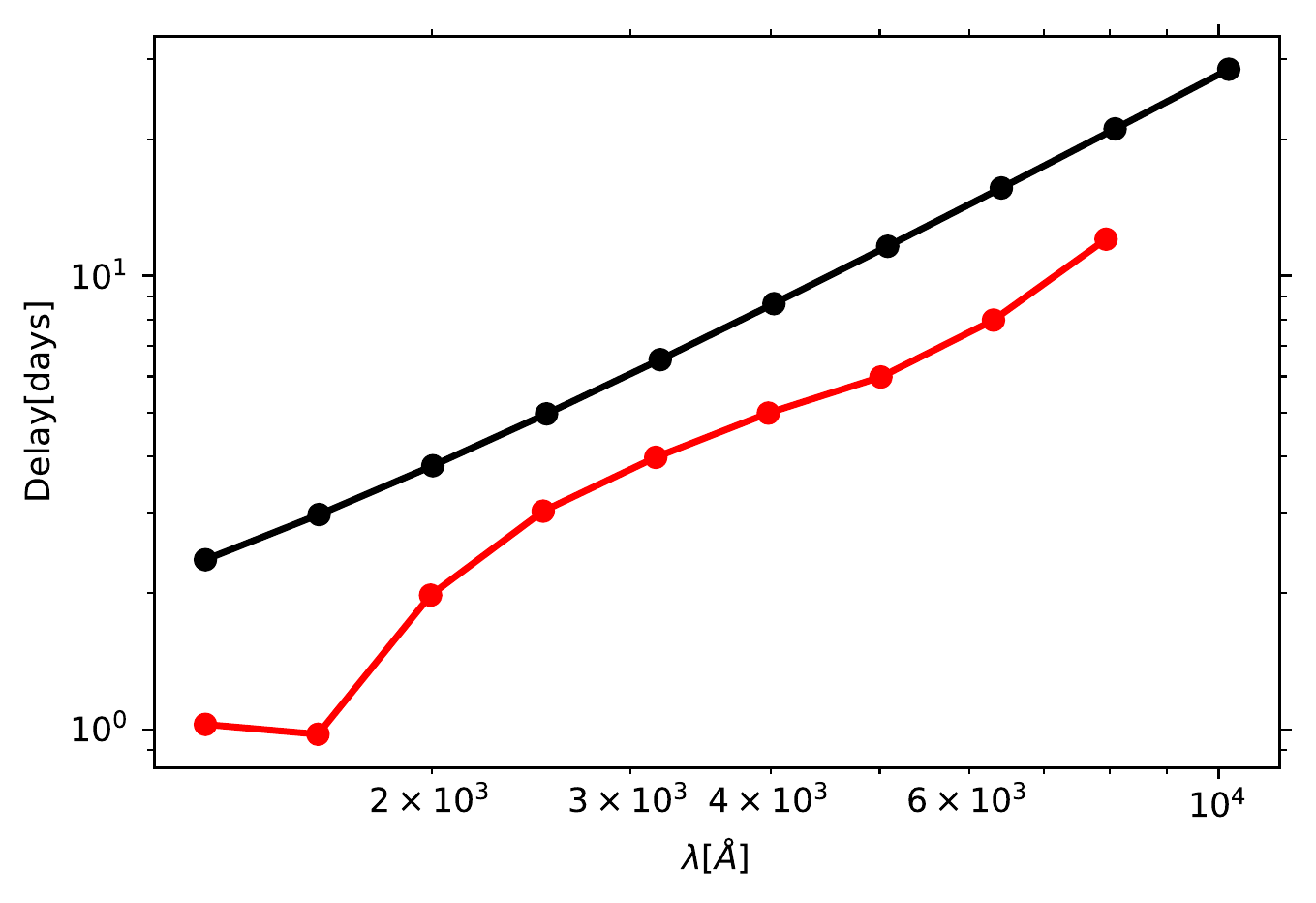}
\caption{\label{fig:delay_disk_curve_big_mass} Time delay from a bare disk as a function of the wavelength calculated with ICCF method from the artificial light curves. Model parameters: Black hole mass $10^9 M_{\odot}$, Eddington ratio = 1.0, corona height = 5r$_g$, inclination angle = 30 degrees, luminosity is of the order of 10$^{45}$ erg s$^{-1}$, and the frequency break corresponding to 75 days.  
}
\end{figure}

 The sampling rate of the lightcurve must be adjusted to the source parameters. Our previous discussion focused on $10^{8} M_{\odot}$ black hole mass. However, if we increase the black hole mass by a factor of 10, the delays are still not well recovered at the shortest wavelengths as shown in Figure~\ref{fig:delay_disk_curve_big_mass}.
However, at longer wavelengths, the delay is comparable to the value expectations based on the response function and the slope is well recovered (see Figure~\ref{fig:delay1}). Thus, for a larger mass, a one-day sampling is fully adequate.

\begin{figure}
\centering
\includegraphics[width=0.5\textwidth]{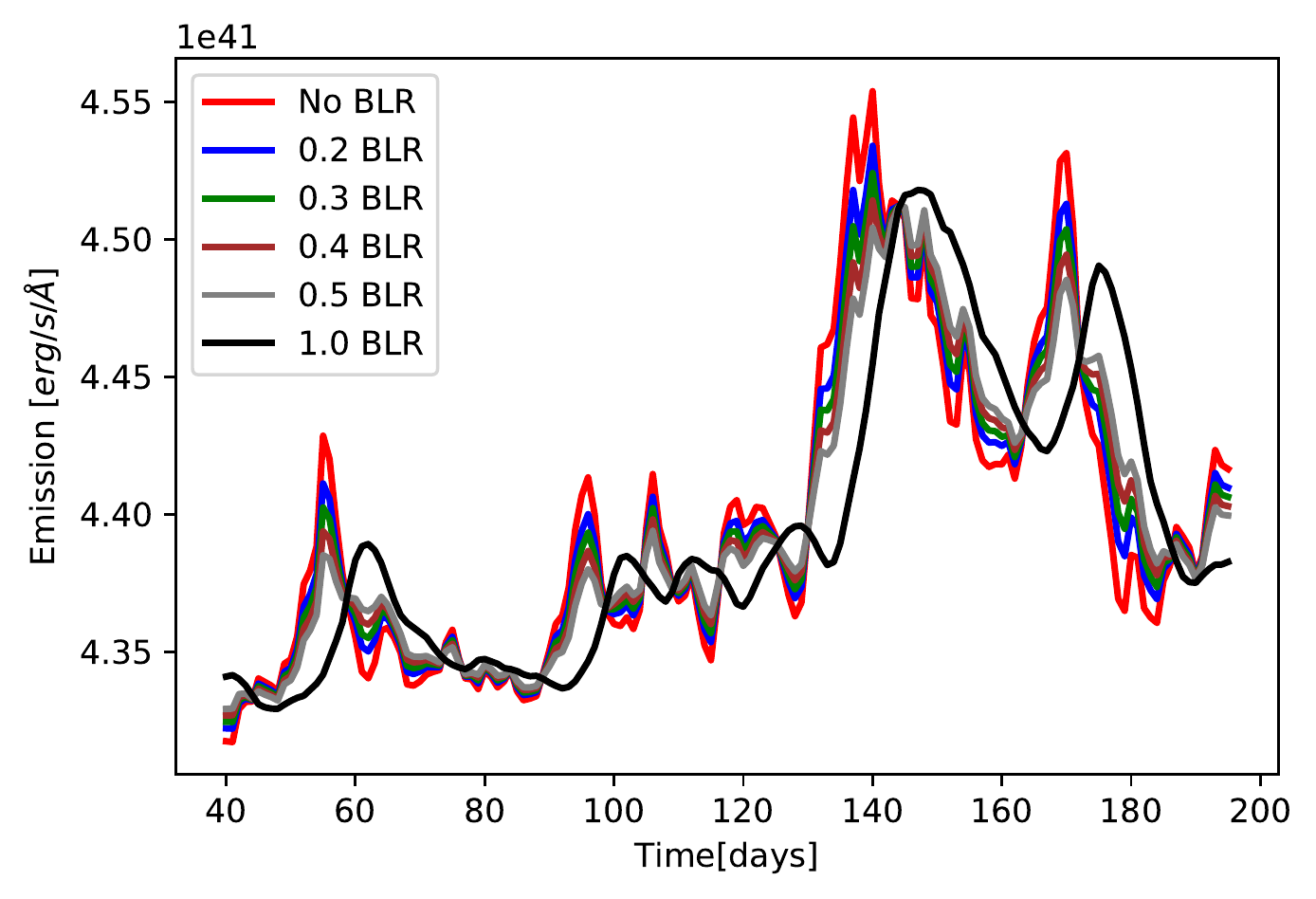}
\caption{\label{fig:disk_light_curve}Plot shows the light curve for $\lambda = 1258.92 \AA$ generated for different BLR contribution. We used half-Gaussian for BLR profile $\sigma = 5$ days and we used an X-ray light curve for breaking time 75 days. Parameters: Black hole mass $10^{8}M_{\odot}$, Eddington ratio = 1.0, height: $h = 5 r_g$, and viewing angle: $i = 30$ degrees.}
\end{figure}

\subsection{The influence of stochastic approach on time delay with BLR contribution}

In the case of the analytical (response function) approach, the presence of the additional scattering in the BLR resulted in a simple shift in the net time delay, as expected previously \citep[e.g.,][]{netzer2021,lawther2018}. However, the stochastic light curve approach may not preserve such a simple trend in all parameter range. We selected the timescale break of 75 days, since it was working relatively well for the disk delay, and we adopted a one-day sampling which was adequate at longer wavelengths.  The resulting light curves thus become distorted (smoothed and shifted) by the BLR (as illustrated in Figure~\ref{fig:disk_light_curve}). Smoothing is clearly stronger when the width of the BLR response function is larger. 
 
 We now calculate the time delay using those stochastic curves. 
 We noticed that the numerically calculated time delay does not increase with $f_{BLR}$ up to the critical moment when $f_{BLR}$ crosses (unrealistic) value of 50\%. This is in contrast to expectations based on the response function approach. The peak in the combined response function is still due to the disk for smaller values of $f_{BLR}$, which is apparently confusing in terms of the numerical method. We see the same effect for the other values of the width of the Gaussian. It may indicate that in real data analysis, we would actually be recovering the disk delay, independently of the BLR contamination.

However, the symmetric Gaussian shape for the BLR response is unlikely, so we repeated the same analysis for half-Gaussian shape.
 We found that for half-Gaussian shape the BLR contamination  shows more similarity between the time delay predicted by response function and the stochastic prediction, although the determined lags are always below the ones expected from the combined response function.  

\begin{figure*}
\centering
\includegraphics[width=0.9\textwidth]{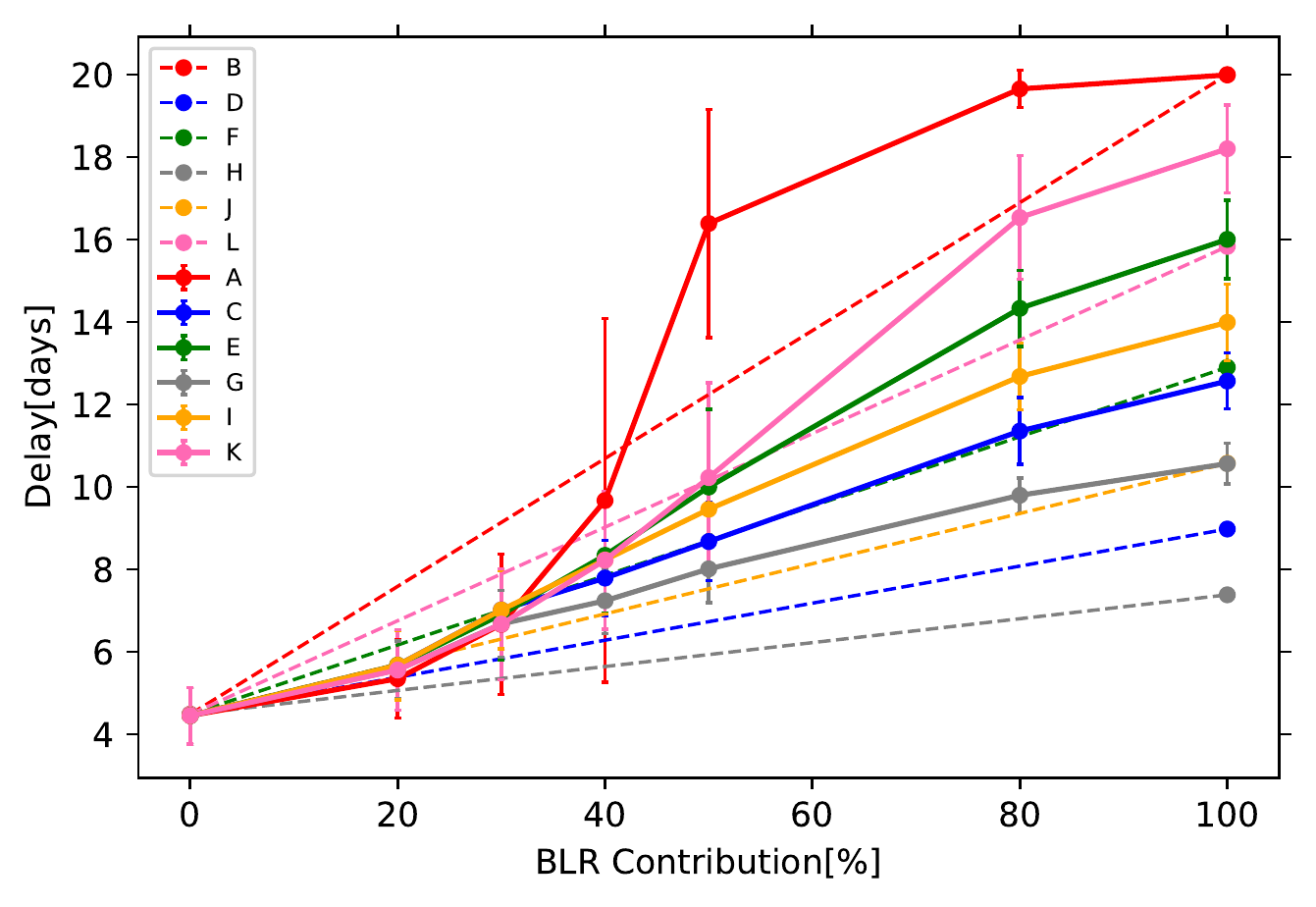}
\caption{\label{fig:netzer} Dashed lines show the linear interpolation between the extreme cases of no BLR contribution to time delay and 100\% of disk photons reprocessed by the BLR. Points show the intermediate delays calculated from stochastic light curves using ICCF between X-ray light curve and  7961.59 $\AA$ for different shapes of the BLR transfer function: A = Gaussian ($\tau_{peak} = 20$, $\sigma$ = 5); C = half-Gaussian ($\tau_{peak} = 5$, $\sigma$ = 5); E = half-Gaussian2 ($\tau_{peak} = 5$, $\sigma$ = 10); G = half-Gaussian3 ($\tau_{peak} = 5$, $\sigma$ = 3); K = half-Gaussian4 ($\tau_{peak} = 5$, $\sigma$ = 7); and I = half-Gaussian5 ($\tau_{peak} = 5$, $\sigma$ = 15). All values of $\tau_{peak}$ and $\sigma$ given in days. Solid lines show the ICCF delay measured between X-ray light curve and  7961.59 $\AA$ light curve by varying the BLR contribution using different BLR response function (B = Gaussian, D = half Gaussian, F = half Gaussian2, H = half Gaussian3, J = half Gaussian4, and L = half Gaussian5).} 
\end{figure*}

 We finally checked, in a systematic way, how the adopted width of the BLR response function affects the delay. This time we performed ten simulations for each parameter set and the errors mark the dispersion.   
We illustrate the complex trend in the time delay with the change of the response model for BLR, $\sigma,$ and $f_{BLR}$ in Figure~\ref{fig:netzer}.

 We see that for a Gaussian shape the departure from the linear trend of the rise of the expected delay with the importance of the BLR contamination is strong. But most of the other shapes also predicted similar trend -- the initial rise was slower than expected and only after crossing rather unrealistic level of BLR contribution (above 50\%), the time delay flipped to values close to the BLR time delay.   For example, when the departure between the measured time delay and the linear time delay is determined at 50\% of the BLR contribution we see a delay longer by 33.88\% (Gaussian), 28.85\% (half-Gaussian), 15.01\% (half-Gaussian2), 35.03\%  (half-Gaussian3), 25.62\% (half-Gaussian4), and 0.66\% (half-Gaussian5). We refer to the caption of Figure~\ref{fig:netzer} for the model parameters. Thus, no departure is seen for very wide asymmetric BLR response profile while narrow asymmetric or symmetric response show a considerable departure from a linear trend.

 We see from the performed simulations that the measured time delay depends on the light curve properties as well as BLR response, and the dispersion in a single measurement is considerable. This means that when an actual reverberation mapping campaign is performed, corresponding to a single realization of our process, some modeling adjusted to the observational setup and source properties is useful for estimating the possibility of the systematic bias in the measured time delays.

 \begin{table*}
 \centering
 \caption{\label{fig:netzer1} Simulated time lags for various shapes of response function and with different BLR contributions. The corresponding plot is shown in Figure \ref{fig:netzer}; $\tau$ is the delay measured in days between X-ray light curve and  7961.59 $\AA$ light curve, and $\Delta \tau$ is error in delay measured in days.}
  \tiny
  \begin{tabular}{ccccccccccccccc}
    \hline
    {$\psi_{BLR}$} &
      \multicolumn{2}{c}{$f_{BLR}=0$ (\%)} &
      \multicolumn{2}{c}{$f_{BLR}=20$ (\%)} &
      \multicolumn{2}{c}{$f_{BLR}=30$ (\%)} &
      \multicolumn{2}{c}{$f_{BLR}=40$ (\%)} &
      \multicolumn{2}{c}{$f_{BLR}=50$ (\%)} &
      \multicolumn{2}{c}{$f_{BLR}=80$ (\%)} &
      \multicolumn{2}{c}{$f_{BLR}=100$ (\%)} \\
      \hline
      & {$\tau$} & {$\Delta \tau$} & {$\tau$} & {$\Delta \tau$} &{$\tau$} & {$\Delta \tau$}& {$\tau$} & {$\Delta \tau$} & {$\tau$} & {$\Delta \tau$} & {$\tau$} & {$\Delta \tau$} & {$\tau$} & {$\Delta \tau$}\\
      \hline
    Gaussian &4.45 &     0.68&
5.34      & 0.95&       
6.66  &      1.70&      
9.67 &       4.40&
16.38&        2.76&
19.65&       0.45&      
19.99&        0.01
 \\
   Half-Gaussian &4.45& 0.68&
5.67&   0.83&
7.01&   0.94&
7.79&   0.91&
8.67&   0.94&
11.3&   0.81&
12.5&   0.67 \\
   Half-Gaussian2 & 4.45&       0.68&
5.67&   0.83&
6.9     &1.1&
8.34&   1.4&
9.99&   1.88&
14.3&   0.92&
16&     0.95 \\
    Half-Gaussian3 & 4.45&      0.68&
5.56&   0.7&
6.67&   0.8&
7.23&   0.79&
8.01&   0.81&
9.79&   0.42&
10.5&   0.49\\
    Half-Gaussian4 &4.45&       0.68&
5.67&   0.83&
7.01&   0.94&
8.22&   1.3&
9.45&   0.83&
12.6&   0.81&
13.9&   0.92 \\
    Half-Gaussian5 & 4.45&      0.68&
5.55&   0.97&
6.67&   1.34&
8.22&   1.67&
10.2&   2.29&
16.5&   1.5&
18.2&   1.06 \\
 \hline
  \end{tabular}
\end{table*}

\section{Discussion}

We studied the wavelength-dependent time delay of optical photons originating from the X-ray photons generated in the lamppost geometry above the AGN accretion disk and reprocessed by the surroundings. We included the photon thermalization and re-emission in the accretion disk, but we also allowed for an additional scattering of the generated optical photons by the inter-cloud medium of the BLR. Such a scattering does not change the photon energy but introduces an additional time delay with respect to the arrival of the primary X-ray emission as well as with respect to the unscattered optical photons. We constructed the response functions for the combined effect of the accretion disk and studied the time delays analytically, but we also constructed simulated X-ray light curves and their reprocessing. 

The results based on the response function computations give a very smooth dependence on the model parameters. The most interesting result of this study is the modification of the time delay by the rising contribution of BLR scattering. This effect is difficult to distinguish from the effect of rising the height of the lamppost, without postulating any contribution from the BLR. In noticing the difference in the curvature of the time delay pattern is practically impossible even with high-quality data, if the mean incident X-ray flux is small in comparison with the disk bolometric luminosity. 
On one hand, this degeneracy between the lamppost height and the BLR contribution can account for surprisingly large heights obtained from data fitting. \citet{kammoun2021_data} successfully modeled the time delay in seven nearby AGNs, but the derived height of the lamppost ranged from  11.2 $R_g$ (for Mkn 509, maximally rotating black hole) to $\sim$ 75 $R_g$ for NGC 7469, independently from the spin. This is not consistent with many of the fits of the X-ray spectra that require low lamppost heights to model the relativistically broadened K$\alpha$ line \citep[e.g.,][]{parker2014,jiang2019,walton2021}. We have an independent insight into the geometry of the X-ray reprocessing from the measurement of the K$\alpha$ line delays, and they rise with the black hole mass from $\sim 100$ to $1000$ s for mass increasing from $10^6 M_{\odot}$ to $10^8 M_{\odot}$ \citep{kara2016}; for  $10^8 M_{\odot}$, this implies a geometrical delay of $\sim 2 R_g$. However, in the case of NGC 7469, the K$\alpha$ line is broad (broadening velocity about 2700 km s$^{-1}$) but not relativistically distorted \citep{mehdipour2015}, so it can come from the outer disk and/or BLR, so a large height is not in contradiction with the X-ray spectrum. 

High values of the irradiating flux allow us to differentiate the delay curve shape caused by the increase of the lamppost height and by the BLR scattering.. The question of whether such high values -- namely, up to 30\% of the disk bolometric luminosity -- are possible is directly related to the question of the origin of the irradiating flux. Hard X-ray emission, as argued by \citet{kubota2018}, contributes less than 2\% to the bolometric luminosity of bright AGNs, not containing inner ADAF. On the other hand, soft X-ray excess can contain much higher fraction of the total flux. The lamppost model is more likely to represent better the hard X-ray emission while the geometry of the soft X-ray emission is still under debate, but it is most likely a warm corona \citep[e.g.,][]{czerny2003,rozanska2015,petrucci2020}. However, studies of other geometries besides the lamppost is beyond the scope of the present paper.

In general, a potential data fitting of the time delay faces a number of degeneracies. As demonstrated by \citet{kammoun2021_data}, independent information about the black hole mass and accretion rate would reduce it considerably; usually, estimates of the black hole mass, based on line widths, are available. Knowing the monochromatic flux, we can also estimate the accretion rate in a way that only weakly depends on the black hole spin. The viewing angle remains, however, an issue, since the monochromatic flux roughly depends on  $\cos(i)$. However, a dusty or molecular torus limits the available viewing angles to between 0 and $\sim70^{\odot}$ \citep[e.g.,][and the references therein]{prince2022}. The new degeneracy between the lamppost height, $H_X$, and the BLR contribution, $f_{BLR}$, creates an additional issue. When the height is small and the high-quality X-ray data are available, we can independently estimate its height, but no such estimate is possible if the height is large and the relativistic distortion of the line  is not strong.

Perhaps, in the future, a more careful modeling of the disk reprocessing plus BLR scattering may help to ease the problem. In our simple code, indeed the effect of the height and the effect of BLR show similar trend for low X-ray luminosity; whereas, in \citet{kammoun2021}, the disk height results in a convex-shape plot of the time delay versus wavelength, while in our simple model the pattern is concave both for height and BLR contribution. We think that the shape should actually be concave, and the convex shape results from too small outer radius adopted in the computations (see Appendix~\ref{appen}). Repeating the calculations of the disk plus BLR scattering using full GR, X-ray reflection, and color correction to the temperature may reveal a systematic difference in the system reaction to these two parameters. In this case, the data fitting should not be done just using a power law part of the delay curve, as in \citet{kammoun2021_data}, but the fits should include the full wavelength-dependent model with the curvature.  Also, studies of the same source at different flux levels are very helpful in disentangling the lamp height and BLR effect, as argued by \citet{vincentelli2022}.

In the present study, we did not include the re-emission by the BLR clouds. Such emission has clear spectral signatures, including the prominent Balmer edge \citep{lawther2018,korista2019,chelouche2019,netzer2021}. This effect is also important but, in principle, it is easier to include it later (in the data fitting), since the prominent Balmer edge should fit the corresponding drop in the time delay. In numerical computations of the reprocessed BLR component with the use of {\sc cloudy} \citep{CLOUDY2017} or equivalent code, the effect of scattering is included but only for (usually) constant density clouds, not accounting for the inter-cloud medium. Thus, the scattering effect can be more difficult to disentangle in the real data. Both broad band data, possibly dense in the wavelength (e.g., coming from specially designed narrow-band filter photometry, as used by \citealt{pozo2019}), but also very dense cadence is essential, as we can see from our experiments with the artificial light curves. Also, the broad  wavelength coverage is very important since it allows to determine the shape of the relation more accurately and to improve the disentangling of the contributions from the disk and BLR.
 Finally, there are two other possible effects that could modify the delay obtained for the disk continuum: the disappearance of the inner cold disk and the disk winds, as argued by \citet{zdziarski2022} -- an insight into this issue could be expected from a fitting of the broadband spectra of the studied objects.

\section{Conclusions}

The results of our modeling of the X-ray reprocessing by the accretion disk, with the additional scattering of disk photons in the BLR region are as follows:
\begin{itemize}
    \item  for low-irradiating X-ray flux, the lamppost height and BLR contribution through scattering are degenerate; 
    \item for high-irradiating flux, there is a difference in the curvature in delay versus wavelength plot that allows us to distinguish between the two effects -- if the wavelength coverage is broad enough;
    \item the time delay rises linearly with the BLR contribution in the description, which uses the response function; 
    \item when stochastic incident light curves are used, the time delay is aptly recovered only if the time-step of the curve is considerably denser than the characteristic variability timescale (set by the high-frequency break in the power spectrum) and when the total duration of the light curve is much longer than this timescale;
    \item in numerical stochastic incident light curves, this linear dependence is perturbed and the time delay rise is initially slower than linear, then rising rapidly with the BLR contribution;
    \item our modeling shows that the results of the time delay based on a single observational campaign should be supplemented with simulations in order to identify the potential bias in measuring the time delays.
\end{itemize}

\section*{Acknowledgements}
We are thankful to Elias Kammoun for helpful discussions,  and to the anonymous referee for the comments which helped considerably to improve the manuscript.
The project was partially supported by the Polish Funding Agency National Science Centre, project 2017/26/\-A/ST9/\-00756 (MAESTRO  9) and MNiSW grant DIR/WK/2018/12. SP acknowledges financial support from the Conselho Nacional de Desenvolvimento Científico e Tecnológico (CNPq) Fellowship (164753/2020-6). This project has received funding from the European Research Council (ERC) under the European Union’s Horizon 2020 research and innovation program (grant agreement No. [951549]). RP and BC acknowledge the Czech-Polish mobility program (M\v{S}MT 8J20PL037
and PPN/BCZ/2019/1/00069).

\bibliographystyle{aa}
\bibliography{main}

\appendix
\section{Comparison of results from our code and from \citet{Kammoun2019}}
\label{appen}

\begin{figure}[htb!]
\centering
\includegraphics[width=0.5\textwidth]{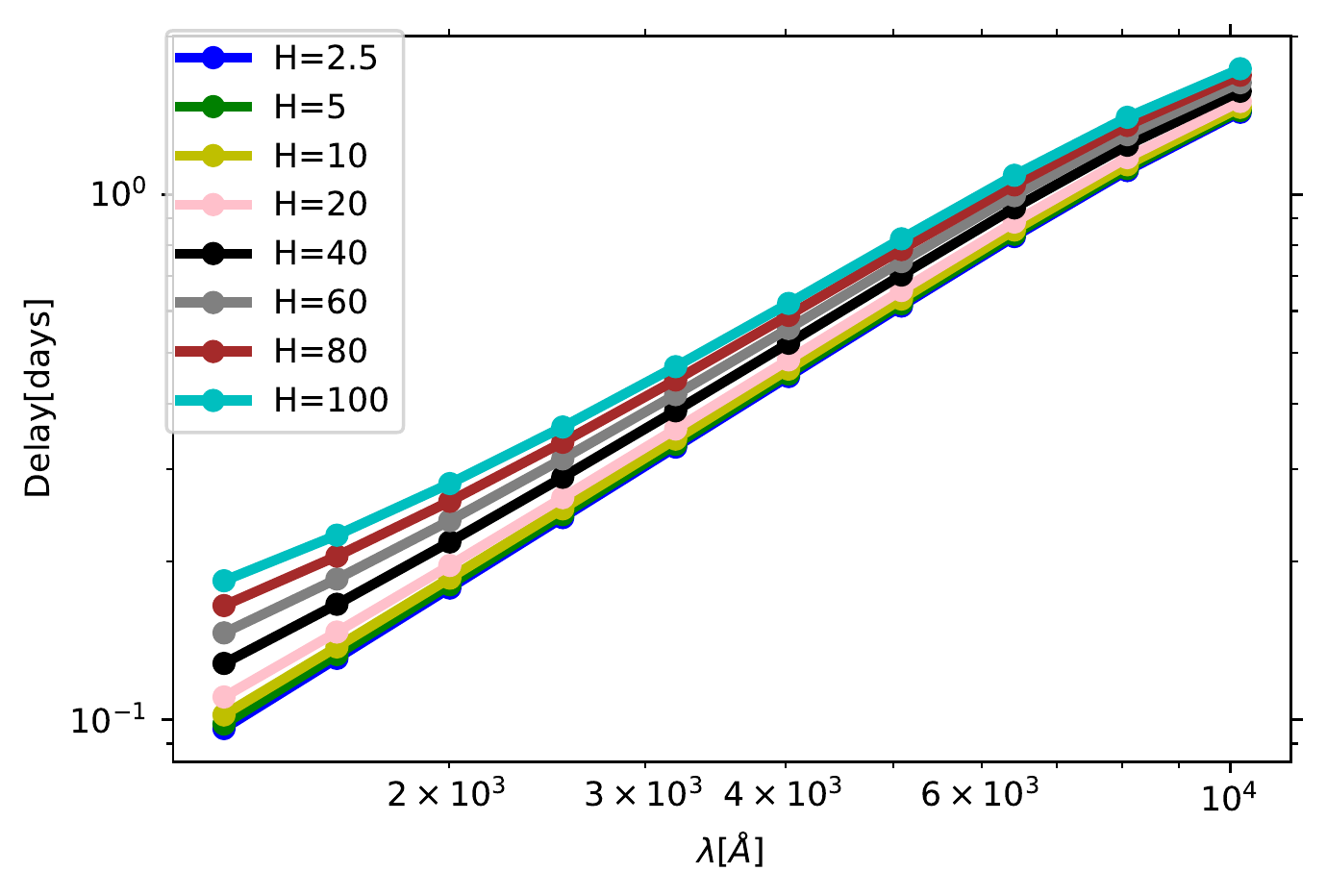}
\includegraphics[width=0.5\textwidth]{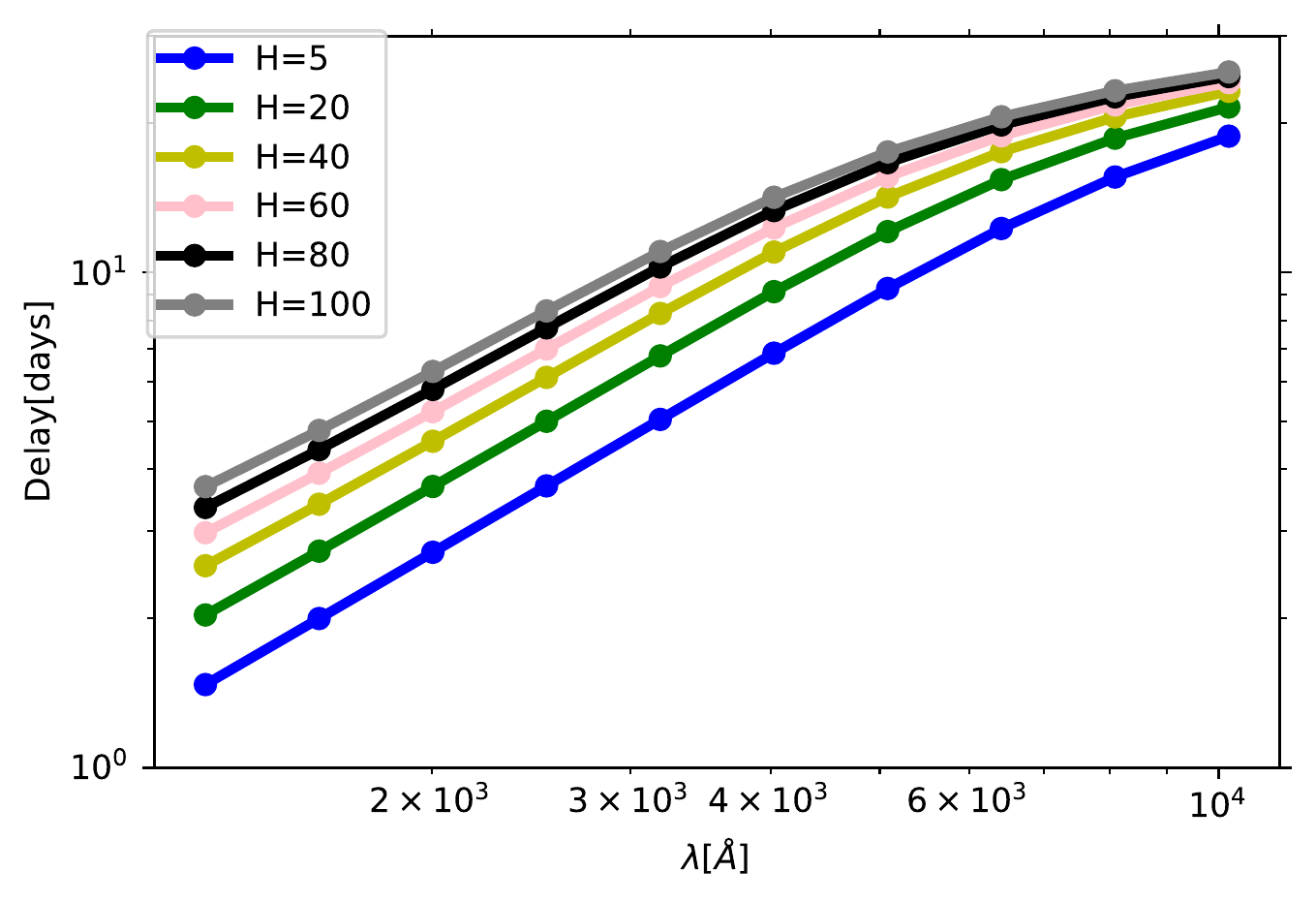}
\includegraphics[width=0.5\textwidth]{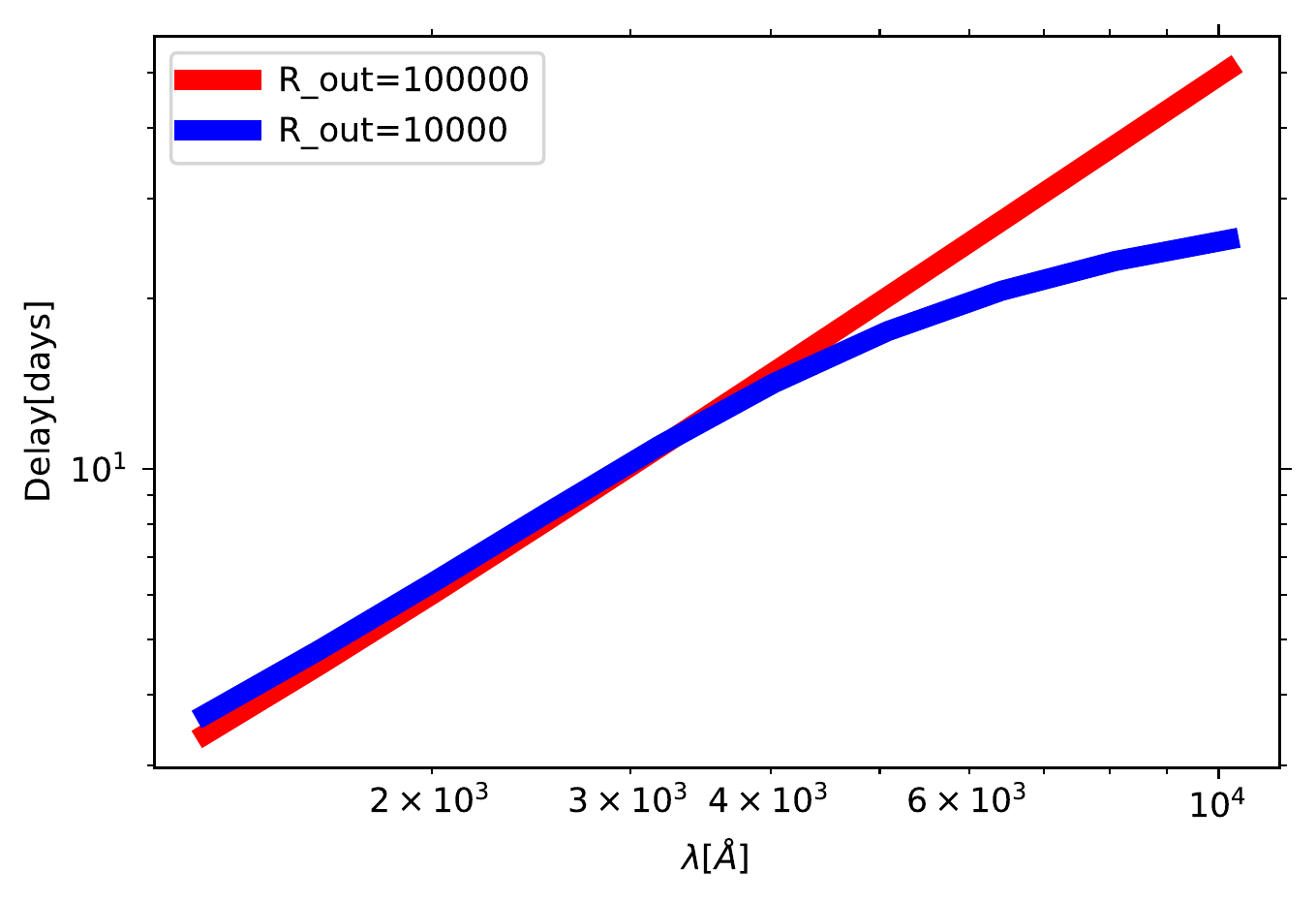}
\caption{\label{fig:elias_comparison} { Comparision of delay plots. Upper panel: Delay curves from our code with the parameters: black hole mass = $10^7 M_{\odot}$, Eddington Ratio = 0.05, inclination angle = 40 degrees, color correction = 2.4, and X-ray source luminosity = $1.26\times10^{43}$ erg s$^{-1}$}. Middle panel: Black hole mass = $10^8 M_{\odot}$, Eddington Ratio = 1.0, inclination angle = 30 degrees, color correction = 2.4, and X-ray luminosity = $3.78 \times10^{46}$ erg s$^{-1}$. Lower panel: Same parameters as middle panel, only for corona height of $100 R_g$, but two different values of the disk outer radius.} 
\end{figure}

To test the importance of the effects neglected in our model, we calculated the model as closely as possible to the standard one of \citet{kammoun2021}. We concentrate on the issue of the height effect on the measured delay. The result from our code is shown in Figure~\ref{fig:elias_comparison}. We use the parameters adopted by \citet{kammoun2021}. We also included  the color correction of 2.4 in this case, unlike in the other plots. We see that our code gives shorter time delays at the shortest wavelengths in comparison with Fig.~18 in \citet{kammoun2021}, since GR effects are most important in the disk central regions. However, at the longest wavelengths, our delays are also somewhat shorter. 
The maximum delay is 1.7 days for $\lambda = 10^4$ \AA~ as shown in Figure~\ref{fig:elias_comparison}, while in \citet{kammoun2021} in their Fig.18 the maximum delay for the same wavelength is approximately 2.8 days.

Since our plot with the color correction shows also traces of the convex shape, we carried out two experiments in order to understand better this trend. We calculated exemplary delay curves for a much higher incident luminosity and in this case, the effect of convex shape is even much stronger (see Figure~\ref{fig:elias_comparison}, middle panel). Since introducing the color correction and increasing the incident flux both lead to an increase in the disk temperature and the emission at a given wavelength comes with increasing disk radii, we checked whether the convex shape is not caused by adopting too small outer radius. Indeed, repeating the computations just for the high luminosity and the lamppost height of 100 $R_g$ for two values of the disk outer radius ($10^4$ and $10^5 R_g$), we show that the convex shape is an artifact of an overly small outer radius value set in the model.

\end{document}